\newcommand\pa{\partial}

\newcommand{\beque}{\begin{equation*}}
\newcommand{\eeq}{\end{equation}}
\newcommand{\beq}{\begin{equation}}
\newcommand{\eeque}{\end{equation*}}
\newcommand{\beqnl}{\begin{eqnarray}}
\newcommand{\eeqna}{\end{eqnarray*}}
\newcommand{\beqna}{\begin{eqnarray*}}
\newcommand{\eeqnl}{\end{eqnarray}}

\providecommand{\abs}[1]{\lvert#1\rvert}

\documentclass[aps,prd,twocolumn]{revtex4}

\usepackage{amsfonts}
\usepackage{amsthm}
\usepackage{amsbsy}
\usepackage{amssymb}
\usepackage{amsmath}
\usepackage{graphicx}
\usepackage{color}

\begin{document}

\date{\today}

\title{Dielectric black holes induced by a refractive index perturbation and the Hawking effect.}

\author{F.~Belgiorno$^{1}$, S.L.~Cacciatori$^{2,3}$, G.~Ortenzi$^{4}$, L.~Rizzi$^{2}$, V.~Gorini$^{2,3}$, D.~Faccio$^{5}$}

\address{$^1$ Dipartimento di Fisica, Universit\`a di Milano, Via Celoria 16, IT-20133 Milano, Italy\\
$^2$ Department of Physics and Mathematics, Universit\`a dell'Insubria, Via Valleggio 11, IT-22100 Como, Italy\\
$^3$ INFN sezione di Milano, via Celoria 16, IT-20133 Milano, Italy\\
$^4$ Dipartimento di Matematica e Applicazioni, Universit\`a di Milano-Bicocca, Via Cozzi 53, IT-20125 Milano, Italy\\
$^5$ School of Engineering and Physical Sciences, SUPA Heriot-Watt University, Edinburgh, Scotland EH14 4AS, UK}


\begin{abstract}

We consider a 4D model for photon production induced by a 
refractive index perturbation  in a dielectric medium. We show that, in this model, we can infer the presence
of a Hawking type effect. This prediction shows up both in the analogue Hawking framework,
which is implemented in the pulse frame and relies on the peculiar properties of the
effective geometry in which quantum fields propagate, as well as in the laboratory frame, through
standard quantum field theory calculations. Effects of optical dispersion are also taken into account, and
are shown to provide a limited energy bandwidth for the emission of Hawking radiation.

\end{abstract}

\maketitle

\section{Introduction}
\label{intro}

One of the most intriguing predictions of quantum fields in curved spacetime geometries is the
production of Hawking radiation. In 1974 S. Hawking predicted that black holes emit particles
with a thermal spectrum.
Therefore a black hole will evaporate,
shedding energy under the form of a blackbody emission \cite{haw-nature,haw-cmp,kiefer}.
However, it turns out that for a typical stellar mass black hole the temperature of this radiation
is so low ($\sim$10 nK) that it has no hope of being directly detected. Notwithstanding,
Hawking radiation does not actually require a true gravitational event horizon but, rather, it
is essentially a kinematical effect, i.e. it requires some basic ``kinematical'' ingredients but no
specific underlying dynamics  (see e.g.
\cite{visser-prl,visser-essential,barcelo-prl}).
What suffices is the formation of a trapping horizon in a curved spacetime metric of any kind,
and the analysis of the behavior of any quantum field therein (see also the discussion in \cite{barcelo-prl}).
The quanta of this field will then be excited according to the prediction of Hawking.
On this basis, a number of analogue systems have been proposed, for the first time by W. Unruh
\cite{unruh} and later by other researchers (see e.g. \cite{barcelo} and references therein),
that aim at reproducing some aspects of the kinematics of gravitational systems. Most of these analogies
rely on acoustic perturbations and on the realization of so-called dumb holes: a liquid or gas medium
is made to flow faster than the velocity of acoustic waves in the same medium so that at the transition
point between sub and supersonic flow, a trapping horizon is formed that may be traversed by the acoustic
quanta, viz. phonons, only in one direction \cite{barcelo}.
Unfortunately the phonon blackbody spectrum is expected to still have very low temperatures,
thus eluding direct detection (see e.g. \cite{macher-bec}). A parallel line of investigation involves effective geometry
for light, which has been introduced by Gordon \cite{gordon} and extended also to
nonlinear electrodynamics;  black hole metrics have been introduced and
the possibility to perform experiments involving Hawking analogue radiation has been explored
\cite{Leonhardt:1999fe,Leonhardt:2000fd,DeLorenci:2001ch,Schutzhold:2001fw,Brevik:2001nf,DeLorenci:2001gf,
Novello:2001gk,Marklund:2001dq,DeLorenci:2002ws,Novello:2002ed,Novello:2003je,Unruh:2003ss,
Schutzhold:2004tv}.
Some time ago Philbin et al. proposed an optical analogue in which a soliton with intensity $I$,
propagating in an optical fibre, generates through the nonlinear Kerr effect a refractive index perturbation (RIP), $\delta n = n_2 I$,
where $n_2$ is the Kerr index \cite{philbin}. The same mechanism has also been generalized to a full 4D geometry by Faccio et al.
\cite{faccio} and has led to the first experimental observation of quanta emitted from an analogue horizon \cite{PRL}. The RIP modifies the
spacetime geometry as seen by co-propagating light rays and, similarly to the acoustic analogy, if the RIP
is locally superluminal, i.e. if it locally travels faster than the phase velocity of light in the medium,
an horizon is formed and Hawking radiation is to be expected.

Here we take into consideration the Hawking effect in dielectric black holes induced by a RIP which
propagates with constant velocity $v$.
We first show that, under suitable conditions, we can re-map the original
Maxwell equations for nonlinear optics into a geometrical description,
in analogy to what is done in the case of acoustic perturbations in
condensed matter. Then we provide 
a model in which the presence of the Hawking radiation can be inferred even
without recourse to the analogous model characterized by a curved spacetime geometry.
Indeed, Hawking radiation is a new phenomenon for nonlinear optics, never foreseen before, which could legitimately
meet a sceptical attitude by the nonlinear optics community, because of a missing
description of the phenomenon by means of the standard tools of quantum electrodynamics.
As a consequence, it is important to recast the analogue picture by
using ``standard'' tools of quantum field theory (without any reference to the geometrical picture).\\

The peculiar feature which distinguishes the dielectric Hawking effect from the traditional
black hole one is the presence of optical dispersion, which gives rise
to relevant physical consequences on the quantum phenomenon of particle creation.
Therefore, we take dispersion into account.
We first introduce a 2D reduction of the model in the presence of dispersion, and see
how the dispersion relation is affected by the frequency-dependence of the refractive index.
Then we discuss both phase velocity horizons and group velocity horizons, which can both play a
relevant role in the physical situation at hand; their qualitative difference and the possibility
to discriminate between them in experiments (numerical and/or laboratory-based) is considered.\\

The structure of the paper is as follows. In section~\ref{sec:model} we introduce our idealized model of a RIP propagating, in a nonlinear Kerr medium, with a constant velocity $v$ in the $x$ direction and infinitely extended  in the transverse $y$ and $z$ directions. In the eikonal approximation, the model is embodied in a suitable wave equation for the generic component of the electric field propagating in the nonlinear medium affected by the RIP. We describe this propagation as taking place in an analogue spacetime metric, written in the pulse frame, where the metric is static and displays two horizons $x_+$ and $x_-$ for propagating photons, analogues to a black hole and a white hole horizons respectively. By assigning to the black hole horizon a surface gravity in a standard way, we calculate the evaporation temperature $T_+$ of the RIP in the latter frame, which turns out to be proportional to the absolute value of the derivative of the RIP's profile, evaluated at $x_+$. We also show that $T_+$ is conformally invariant, as expected for the consistency of the model.

In section~\ref{sec:QFT}, we tackle the model from a different perspective, namely within the framework of quantum field theory. We construct a complete set of positive frequency solutions $\Phi_{{\mathbf{k}_l}}$ of the wave equation in the lab frame, in terms of which we quantize the field in the standard way. Then, we compare the $\Phi_{{\mathbf{k}_l}}$ with the corresponding plane wave solutions of the wave equation in the absence of the RIP. This enables us to calculate the Bogoliubov coefficients, in terms of which we express the expectation value of the number operator of the outgoing quanta in the \emph{in}  vacuum state. As expected, this average is a thermal-like distribution displaying the typical $\cos\theta$ temperature profile in the lab frame, where $\theta$ is the emission angle with respect to the direction of motion of the RIP. We then display the transformation law which relates the temperatures in the pulse and in the lab frame, respectively, by means
 of the usual Doppler formula.

The results worked out in the preceding sections apply in the approximation in which dispersion is neglected. In section~\ref{sec:comment} we study the modifications we expect when dispersion is taken into account.
In particular, we show that, whereas in the dispersionless case one should expect to observe all the blackbody spectrum, in the presence of dispersion only a limited spectral region, which depends on $v$, will be excited, and the blackbody spectrum shape will not be discernible. Moreover, one can introduce two different concepts of
horizon: phase velocity horizons and group velocity horizons, which are shown to be involved with different
spectral bandwidths, and with very different qualitative behavior regarding their action on photons.

Section \ref{sec:conclusion} is devoted to the conclusions.

Finally, we have added three appendices. The first establishes a correspondence between our black hole metric and the acoustic one. The second one relates the temperature to the standard conical singularity of the Euclidean version of the metric. The third establishes the relevant analytic properties of the Bogoliubov coefficients.

\section{Static dielectric black hole}\label{sec:model}

We consider a model-equation which is derived from
nonlinear electrodynamics (with $\chi_2=0$ and $\chi_3\ne 0$) in the eikonal approximation for a perturbation of a
full-nonlinear electric field propagating in a nonlinear Kerr medium. In order to make the forthcoming analysis as simple as possible
we replace the electric field with a scalar field $\Phi$ (cf. e.g. Schwinger's analysis of sonoluminescence \cite{schwinger}), for which we obtain
the wave equation \cite{faccio-spie}
\beq
\frac{n^2 (x_l-vt_l)}{c^2} \partial_{t_l}^2 \Phi - \partial_{x_l}^2 \Phi- \partial_y^2 \Phi  - \partial_z^2 \Phi=0.
\label{model}
\eeq
Coordinates in the lab frame are denoted by $t_l,x_l,y,z$ (the labels of
$y,z$ are omitted because they will not be involved in the boost relating the
lab frame with the pulse frame).
Here, $n(x_l-vt_l)$ is the refractive index, which accounts for the propagating RIP in the dielectric. The latter can be obtained by means of an
intense laser pulse in a nonlinear dielectric medium (Kerr effect).
In our model the RIP does not depend on the transverse coordinates, namely it is infinitely extended in the transverse dimensions.
This approximation is necessary in order to carry out the calculations below, which allow us to draw a tight analogy between the Hawking geometrical
description and the quantum field theoretical treatment. Such an infinitely extended RIP is clearly an idealization, which cannot be obtained in
an actual experiment.
However, relatively flat,
e.g. super-Gaussian-like pulses may be produced that would, at least locally, fall within the approximations adopted here.

Eq. (\ref{model}) arises also in the eikonal approximation for a scalar field in the metric
\beq\label{metric}
ds^2 = \frac{c^2}{n^2 (x_l-vt_l)} dt_l^2 - dx_l^2-dy^2-dz^2.
\eeq

To carry out the analysis in the context of the analogue metric, we pass from the laboratory frame to the pulse frame
(refractive index perturbation frame) by means of a boost: $t = \gamma (t_l-\frac{v}{c^2} x_l)$,
$x = \gamma (x_l-v t_l)$, and we obtain
\begin{multline}
ds^2 = c^2 \gamma^2 \frac{1}{n^2} (1+\frac{n v}{c})(1-\frac{n v}{c}) dt^2 + \\
+ 2 \gamma^2 \frac{v}{n^2} (1-n^2) dt dx -\gamma^2 (1+\frac{v}{n c})(1-\frac{v}{n c}) dx^2-\\ -dy^2-dz^2.
\end{multline}
We assume
\beq \label{nx}
n(x)=n_0+\delta n= n_0+\eta \bar{I}(x),
\eeq
\begin{figure}[t]
\includegraphics[angle=0,width=8cm]{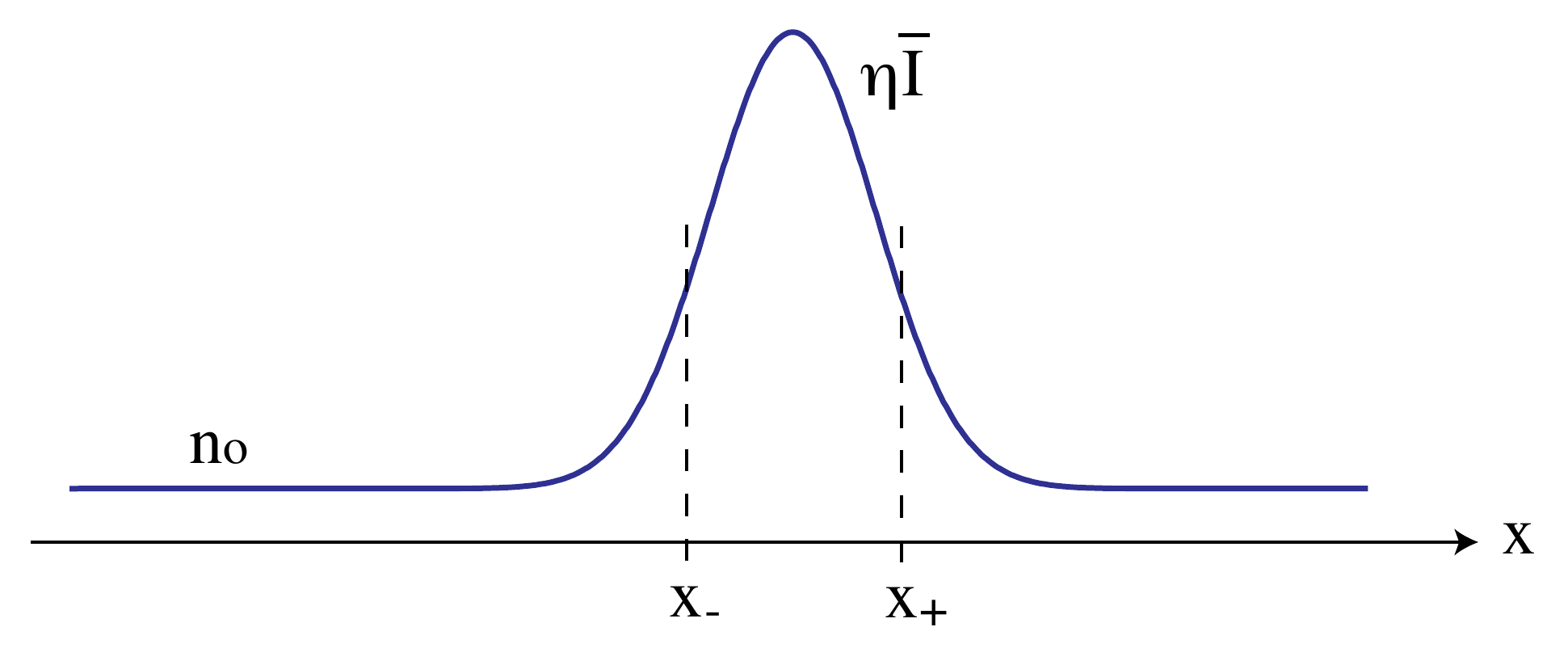}
	\caption{ \label{fig:fig1a} (in color online) Scheme of the RIP geometry. $x_+$ and $x_-$ indicate the black hole and white hole horizon positions, respectively.}
\end{figure}
\noindent
where $\eta$ is meant to be positive, with $\eta \ll 1$ due to the actual
smallness of the Kerr index
(a negative $\eta$ could be easily taken into
account); $\bar{I}$ denotes the normalized intensity of the pulse, is taken to be a $C^{\infty}$ function, rapidly decaying at infinity and
with a single maximum at $x=0$, of height $1$. A scheme of the RIP is shown in Fig.~\ref{fig:fig1a}.
The form of $\bar{I}$ implies that both $\partial_t=:\xi$ and $\partial_{\phi}$ are Killing vectors for
the given metric.
Then the surface $g_{00}=0$ is lightlike and corresponds to an event horizon.
It is also possible to verify that the Frobenius integrability conditions are trivially
satisfied, so that there exists a coordinate transformation carrying the metric
into a static form (see e.g. \cite{wald}). Even if these coordinates are singular, we
carry out the relative transformation because, on one hand, it allows a more straightforward
comparison with the well-known Schwarzschild case, and on the other hand, it allows a
direct computation of the greybody coefficient (cf. sect. \ref{sec:greybody}).
To implement this transformation consider the following coordinate change:
\beq
dt = d\tau -\alpha (x) dx,
\eeq
where
\beq
\alpha (x) = \frac{g_{01} (x)}{g_{00} (x)}.
\eeq
Then the metric takes the static form
\beq\label{static-pulse}
ds^2 = \frac{c^2}{n^2 (x)} g_{\tau \tau} (x) d\tau^2 - \frac{1}{g_{\tau \tau} (x)} dx^2 -dy^2-dz^2,
\eeq
where
\beq\label{pulse-lapse}
g_{\tau \tau} (x) : = \gamma^2 \left(1+n(x)\frac{v}{c}\right)\left(1-n(x) \frac{v}{c}\right).
\eeq
There is a remarkable resemblance of the
$\tau,x$-part of the metric with a standard static spherically symmetric metric in general relativity in the so-called
Schwarzschild gauge, aside from the important difference represented by the factor $\frac{c^2}{n^2}$
replacing $c^2$. The horizons are determined by the condition $g_{\tau \tau} =0$, i.e. by
\beq
\label{horizons1D}
1-n(x)\cfrac{v}{c}=0,
\eeq
which, by Eq.~(\ref{nx}), becomes $n_0+\eta \bar{I}(x)=c/v$. Due to the specific shape assumed for $\bar{I}$ there will be two horizons:
one, denoted by $x_+$, located on the rising edge of the RIP (i.e. $dn/dx|_{x_+}<0$) and one, denoted by $x_-$, on the falling edge
(i.e. $dn/dx|_{x_-}>0$). Since $0\leq \bar{I}\leq 1$, the condition for the occurrence of the event horizons is
\beq
\frac{1}{n_0+\eta} \leq \frac{v}{c} < \frac{1}{n_0}.
\label{cond-ergo-eta}
\eeq
The external region corresponds to $x<x_-$ and to $x>x_+$. The leading horizon $x=x_+$ is
a black hole horizon, whereas the trailing one $x=x_-$ is a white hole horizon. This can be understood as follows.
Consider the front observer, i.e. the observer toward which the pulse is moving.
Any photon starting from the region behind the front zero $x_+$ cannot reach this observer before the latter is reached by the pulse.
Indeed, the pulse is moving superluminally relative to any photon travelling in the region between the two zeros $x_-$ and $x_+$.
Therefore, the ``leading'' zero $x_+$ corresponds to a black hole horizon. Analogously, the ``trailing'' zero $x_-$ corresponds to a white
hole horizon. See also Appendix \ref{acou} for further details.

\subsection{Surface gravity and temperature}
We can formally assign a temperature to the ``black hole'' horizon $x_+$ by defining it in the usual way in terms of
a ``surface gravity'' $\kappa_+$ associated to the latter.
We have
\beqnl
\kappa_+^2 &:=& -c^4 \frac 12 g^{ab} g^{cd} (\nabla_a \xi_c) (\nabla_b \xi_d)|_{x=x_+}= \cr
& -&\frac{c^4}{2}\left[  \frac{-2 \gamma^4}{n^4} \left(\frac{dn}{dx}\right)^2  \right]_{x=x_+},\label{prima}
\eeqnl
or \footnote{the factor $c^4$ appears since we are using standard units in place of the customary natural ones}
\beq \label{surf_grav}
\kappa_+ =  \gamma^2  v^2 \left|\frac{dn}{dx}(x_+)\right|.
\eeq
Then, the temperature is given by the familiar formula which has been also derived in
\cite{philbin}:	
\beq \label{Hawking_temp}
T_+ = \frac{\kappa_+ \hbar}{2\pi k_b c}=\gamma^2  v^2 \frac{\hbar}{2\pi k_b c}
\left| \frac{dn}{dx}(x_+)\right|.
\eeq
We can arrive at formula (\ref{surf_grav}) by several different methods. For the sake of
completeness, in Appendix \ref{temp-confo} we employ a different method to derive the expression (\ref{Hawking_temp}) for the temperature, and its invariance
with respect to the conformal factor. The latter aspect is well-known in General Relativity \cite{kang},
and also
in the frame of acoustic black holes \cite{visser-acou}.\\
Since
\beq
\frac cv =n (x_+) = n_0 + k \eta,
\eeq
with $k=\bar{I}(x_+) \in (0,1)$, we can rewrite equations (\ref{surf_grav}) and (\ref{Hawking_temp}) respectively as follows:
\begin{gather}
 \kappa_+ = \frac{c^2}{(n_0 + k \eta)^2-1} \left|\frac{dn}{dx}(x_+)\right|,\\
T_+ =\frac{ \hbar c}{2\pi k_b } \frac{1}{(n_0 + k \eta)^2-1} \left|\frac{dn}{dx}(x_+)\right|.
\end{gather}

\subsection{Gaussian pulse}
As an example of a refractive index perturbation $\eta \bar{I}(x)$ consider a Gaussian normalized intensity
of the pulse,
\beq\label{gauss}
\bar{I} (x) = \exp \left( - \frac{x^2}{2 \sigma^2} \right)
\eeq
and  $\eta$ a small parameter. These choice reflects a typical situation in experimental optics.
Then the horizons given by Eq.~(\ref{horizons1D}) ) occur for
\beq
x_{\pm} = \pm \sigma \sqrt{ - 2 \log \left[ \left( \frac{c}{v} - n_0 \right) \frac{1}{\eta} \right] },
\eeq
and $x_- = -x_+$.
The surface gravity and the temperature are respectively given by
\beq
\kappa_+ = \gamma^2 v^2 \frac{k \eta}{\sigma}  \sqrt{2 \log \frac{1}{k}},
\eeq
and by
\beq
T_+ = \frac{\kappa_+ \hbar}{2\pi k_b c}\sim \frac{\hbar c}{2 \pi k_b} \frac{1}{n_0^2-1} \frac{k\eta}{\sigma} \sqrt{2\log \frac{1}{k} }.
\eeq
Typical values of the parameters are:
$\sigma \sim 10^{-5} m,\eta\sim 10^{-3}, n_0\sim 1.45$.
Then, for $k=\frac 12$, say, we get
\beq
T_+\sim 2\cdot 10^{-2} K,
\label{t-pulse}
\eeq
which is greater than in standard cases of sonic black holes. This temperature value is not to be meant as
characteristic, somewhat larger values may be obtained. See the following
subsection.\\
Note that this is the temperature in the frame of the pulse,
with which the geometry is associated.
A further transformation is needed in order to recover the temperature in the laboratory frame (see the following
section).
Note also that the temperature is proportional to the derivative with respect to $x$ of the refractive
index, evaluated at the black hole horizon. This is in agreement with the results of Ref.~\cite{philbin}.

\subsection{Shockwave model}
\label{sec-shock}

Let us assume that the rear part (trailing edge) of the filament is characterized by a shockwave profile
(see e.g. \cite{philbin}). This is a typical e.g. of  spontaneous laser pulse dynamics in transparent media with third order nonlinearity.
An analogous behavior is expected for the refractive index. As a
consequence, in our model it is necessary to introduce, beyond the scale $2\sigma$, which, roughly
speaking, represents the overall spatial extension of the RIP, at least a further scale that
corresponds to the `thickness' of the region where the refractive index undergoes its most
rapid variation. Then we can adopt a profile which is analytically described as follows:
let us define
\beq
H(x):= 1+ \tanh \left(\frac{\sigma+x}{\delta_{wh}}\right)
\tanh \left(\frac{\sigma-x}{\delta_{bh}}\right),
\eeq
where $\delta_{wh},\delta_{bh}>0$ are length scales describing the `thikness' of the region over
which a rapid variation of the refractive index occurs, and $2\sigma>0$ corresponds to the overall
spatial extension of the RIP. Compare also the choice of the velocity profile for acoustic geometries
chosen in \cite{coutant-laser}.
Then for the refractive index profile we choose
\beq
n(x)=n_0+\eta \frac{H(x)}{\max_x H(x)},
\eeq
It is evident that asymptotically $n$ converges to $n_0$ and the RIP is correctly normalized.
This model can be also useful in view of exploring the possibility to obtain a bh-wh laser  \cite{coutant-laser}.\\
From a physical point of view, we point out that, in the presence of a shock front
typical values of $\delta_{wh},\eta$ may be order of $\delta_{wh}\sim 10^{-6}$m and $\eta\sim 5\cdot 10^{-3}$
respectively, leading to $T_+\sim 1$ K.\\

\section{Quantum field theoretical treatment: Hawking radiation in the laboratory frame}\label{sec:QFT}
The deduction of Hawking radiation emission by a (non-extremal)
black hole can also be carried out in a dynamical situation where the effects of the onset
of a black hole horizon on quantum field theory modes is taken into account. See e.g. the seminal paper
by Hawking \cite{haw-cmp} and also \cite{unruh-prd}, where the equivalence between the static and the
dynamical picture for Hawking radiation is considered. In the following section, we develop a
dynamical picture for particle production, with the aim of not relying only on the analogue gravity
picture but also obtaining the phenomenon in a non geometrical setting by means of standard tools
of quantum field theory. \\
Consider a massless scalar field propagating in a dielectric medium and
satisfying Eq.~\eqref{model}.
It is not difficult to show that this equation arises in the eikonal approximation
for the components of the electric field perturbation in a nonlinear Kerr medium \cite{DeLorenci:2001ch,Leonhardt:2000fd}.
$n(x_l-vt_l)$ is the refractive index, which accounts for the propagating refractive index perturbation in the dielectric.

In order to develop the model we adopt the following strategy: a) we look first for
a complete set of solutions of equation (\ref{model});
b) then we employ these solutions to perform a comparison between an initial situation, consisting of an unperturbed
dielectric without a laser pulse, with a uniform and constant refractive index $n_0$,
and a final situation where a laser pulse inducing a superluminal RIP is instead present.
This scenario is analogous to the original situation envisaged and treated by Hawking \cite{haw-cmp} in which one considers an
initial spherically symmetric astrophysical object, with no particles present at infinity, followed by a collapsing phase
leading eventually to an evaporating Schwarzschild black hole. In analogy with the original treatment by Hawking,
our aim is to evaluate the mean value of the number of quanta (calculated by means
of final creation and annihilation operators) on the initial vacuum state.

\subsection{\emph{Out} states}

Recast equation (\ref{model}) in terms of the new retarded and advanced variables, respectively $u=x_l-vt_l$ and $w=x_l+vt_l$:
\begin{multline}\label{equation-uandv}
\frac{n^2(u)v^2}{c^2}\left(\pa_u^2\Phi+\pa_w^2\Phi -2\pa_u\pa_w\Phi\right) - \\
- \left(\pa_u^2\Phi+\pa_w^2\Phi +2\pa_u\pa_w\Phi\right) -
\partial_y^2 \Phi - \partial_z^2 \Phi = 0,
\end{multline}
and look for monochromatic solutions of the form
\beq\label{monochr}
\Phi (u,w,y,z) = A(u) {\mathrm e}^{i k_w w + i k_y y + i k_z z}.
\eeq
Then, the ansatz of Eq.~\eqref{monochr} leads to
\begin{multline}\label{eqnforA-4d}
A^{\prime\prime}(u) + 2ik_w \frac{c^2+n^2(u)v^2}{c^2-n^2(u)v^2} A^\prime(u)- \\ - \left(
k_w^2+\frac{k_y^2+k_z^2}{1-n^2(u) \frac{v^2}{c^2}}\right) A(u) = 0.
\end{multline}
Assuming $n(u)$ to be analytic (see e.g.(\ref{gauss})) we see that the coefficients of $A'(u)$ and of $A(u)$
have a first order pole at the roots of
\beq
1-n(u)\frac{v}{c} = 0.
\label{hor}
\eeq
Then, assuming $n$ to be of the form (\ref{nx}), the condition for the occurrence of the event horizons is given by
(\ref{cond-ergo-eta}). Since $x=\gamma u$, the black hole and white hole horizons, when they exist, are located respectively at
\beq
u_\pm=\frac 1\gamma x_\pm,
\eeq
and we have a second order linear differential equation with Fuchsian singular points at $u=u_\pm$ (with $u_-<u_+$).
The general solution of Eq.(\ref{equation-uandv}) can be determined, in the neighbourhood of the singular
points $u_\pm$, by the standard methods of integration by series. We perform this calculation in Appendix \ref{exact-pole}. Here,
instead, since we are interested in the asymptotic behaviour of the solutions as $u\to u_\pm$ and as $u\to \pm \infty$, the
best strategy is to appeal to the WKB approximation. Namely, we write $\Phi$ as in Eq.(\ref{monochr}) with
\beq\label{eik}
A(u) = \exp \left( i \int^u k_u (u') du'\right).
\eeq
Then, we find the following dispersion relation:
\beq\label{eikdisp}
\frac{n^2 v^2}{c^2} (k_u  - k_w)^2 - (k_u  + k_w)^2 -k_{\perp}^2 =0,
\eeq
where $k_\perp^2 = k_y^2 + k_z^2$.
By solving for $k_u $ as a function of $k_w,{\mathbf{k}}_{\perp}$ one arrives at a second degree equation
whose solutions are of the form:
\beqnl
k_u^\pm &=& -\frac{k_w}{1-\frac{n^2 v^2}{c^2}} \left[ 1+ \frac{n^2 v^2}{c^2} \pm \right. \cr
&& \left. \pm 2 \frac{n v}{c} \sqrt{1-\frac{k_{\perp}^2}{k_w^2}
\frac{1-\frac{n^2 v^2}{c^2}}{4 \frac{n^2 v^2}{c^2}}} \right].
\eeqnl
The constraint
\beq
\frac{k_{\perp}^2}{k_w^2}\leq \frac{4 \frac{n^2 v^2}{c^2}}{1-\frac{n^2 v^2}{c^2}},
\eeq
must hold in order that propagating solutions are available. This seemingly would imply
the existence of a limiting angle for the emission, but one has to take into consideration
that, in approaching the horizon, the right hand side of the latest equationn becomes infinite,
leaving room for no real limitation on the emission angle.

The root $k_u^+$ is singular at the horizons, whereas $k_u^-$ is regular there.
We focus on the behaviour of $k_u^+$ in the neighbourhood of $u_+$. Near $u_+$ the dependence on ${\mathbf{k}}_{\perp}$
is washed out, in agreement with the analysis in \cite{visser-essential}, and we have
\beq
k_u \sim -\frac{2 k_w }{1-n \frac{v}{c}}.
\eeq
Then,
\beq
A(u) \sim \exp\left[i \frac{2 c k_w}{v n'(u_+)} \log (u-u_+)\right].
\label{sing-log}
\eeq
The logarithmic divergence in the phase as $u \downarrow u_+$
is entirely analogous to the one experienced by the covariant wave equation at the horizon
of a Schwarzschild black hole \cite{haw-cmp,kiefer} and represents the typical behaviour
of outgoing modes approaching the horizon \cite{visser-essential}.
It is worth to remark that this divergence appears here without any reference to geometry. In addition, as shown in Appendix
\ref{exact-pole}, it emerges naturally as an exact asymptotic (as $u\downarrow u_+$) consequence of Eq. (\ref{equation-uandv}).
On the other hand, the same asymptotic behaviour can be inferred also in the context of the geometric approach. Indeed,
such behaviour is characteristic also of the monochromatic solutions of the covariant wave equation in the metric
(\ref{metric}), though the latter equation is not exactly solvable.

Regarding the dependence on $\vec{k}_{\perp}$, we note that it does not disappear far away from the horizon,
but this turns out not to be relevant for the Hawking phenomenon \cite{visser-essential}.
Also note that $k_u$ has the opposite sign of $k_w$, and $\abs{k_u } > \abs{k_w}$.

We denote by $F^+$ the solutions associated to the root $k_u^+$.
Their asymptotic form (at
large distances from the pulse, i.e. for $u\gg u_+$) is
\beq
F^+ \sim e^{i k_u  u + i k_w w + i k_y y + i k_z z},
\eeq
or
\beq
F^+ \sim e^{i (k_u  + k_w) x_l + i k_y y + i k_z z - i v t_l(k_u  - k_w)}.
\eeq
This can be written as
\beq\label{asympt}
F^+= F^+_{{\mathbf{k}_l}} \sim  \exp\left(i {\mathbf{k}_l}\cdot\mathbf{x}_l  - i\omega_l t_l \right),
\eeq
where ${\mathbf{k}_l}=(k_{xl},k_y,k_z)$,
with
\beqnl
\omega_l &=& v (k_u  - k_w),\label{disp-rel-om}\\
k_{xl} &=& k_u  + k_w,\label{disp-rel-kx}
\eeqnl
so that
\beqnl
k_u  &=& \frac 12 \left( k_{xl}  + \frac{\omega_l}{v}\right),\label{disp-ku}\\
k_w  &=& \frac 12 \left( k_{xl}  - \frac{\omega_l}{v}\right).\label{disp-kw}
\eeqnl
We look for solutions which are asymptotically of positive frequency and outgoing with
respect to the dielectric perturbation, which is right moving. In other words, we choose
$\omega_l >0$ and $k_{xl} >0$. Moreover,
for definiteness we fix $k_w<0$. As a consequence, \eqref{disp-rel-om} and \eqref{disp-rel-kx}
are satisfied for $k_u>-k_w$.
As to $\omega_l$, it satisfies the
asymptotic dispersion relation $n^2_0 \omega_l^2 = {\mathbf{k}_l}^2 c^2$.
A similar analysis can be performed for the solution which is regular at $u=u_+$, which we denote by
$F^+_{{\mathbf{k}_l}, reg}$.
Similarly, we denote by $F^-_{{\mathbf{k}_l}}$ and by $F^-_{{\mathbf{k}_l},reg}$ the singular and, respectively, regular solutions at $u_-$.

Introducing the angle of emission $\theta$, defined by $k_x=|{\mathbf{k}_l}| \cos(\theta)$, Eq. (\ref{disp-kw}) can be written as
\beq
k_w = - \frac{\omega_l}{2 v} \left(1- \frac vc n_0 \cos(\theta)\right).
\eeq

\subsection{\emph{In} states}
Prior to the formation of the pulse the wave equation is
\beq
\frac{n_0^2}{c^2} \partial_{t_l}^2 \Phi - \partial_{x_l}^2 \Phi - \partial_y^2 \Phi  - \partial_z^2 \Phi=0,
\eeq
whose independent positive frequency plane wave solutions are trivially given by (with $\omega_l>0$),
\beq
F^\text{in}_{{\mathbf{k}_l}} =\exp \left(  i {\mathbf{k}_l} \cdot \mathbf{x}_l - i \omega_l t_l \right),
\eeq
where the obvious dispersion relation holds:
\beq
\frac{n_0^2}{c^2} \omega_l^2-k_{xl}^{2}-k_{\perp}^{2}=0.
\eeq
and where $k_\perp^{2} = k_y^{2} + k_z^{2}$.

\subsection{Quantization}
We can separate the monochromatic solutions $\{F^+,F^+_{reg}\}$ corresponding to all possible values of ${\mathbf{k}_l}$, into positive and negative frequency components $\{f^{\phantom{\star}}_{\mathbf{k}_l},f_{\mathbf{k}_l}^\star\}$. These form a complete set of solutions which are mutually orthogonal w.r.t. the index ${\mathbf{k}_l}$ relative to the Klein-Gordon product in the retarded variables:
\beq\label{kgprod}
(\Psi,\Phi) = i \int_{\mathbb{R}^3} du dy dz \left( \Psi^\ast \frac{\partial \Phi}{\partial w} - \Phi \frac{\partial \Psi^\ast}{\partial w}\right).
\eeq
The above product is a function of $w$
(due to the factor $n^2(x_l -v t_l)$ it is indeed impossible to define in a standard way a conserved inner product for Eq.~\eqref{model}).
Expanding an arbitrary solution $\Phi$ of Eq.~\eqref{model} over the modes
$\{f^{\phantom{\star}}_{\mathbf{k}_l},f_{\mathbf{k}_l}^\star \}$ we quantize the field by promoting $\Phi$ to a field operator according to the expansion
\beq
\Phi = \int d^3 {\mathbf{k}_l} \left(a^{\phantom{\star}}_{{\mathbf{k}_l}}f_{\mathbf{k}_l} + a^\star_{{\mathbf{k}_l}}f_{\mathbf{k}_l}^\star\right),
\eeq
where the creation and annihilation operators $a^{\phantom{\star}}_{{\mathbf{k}_l}}$, $a_{\mathbf{k}_l}^\star$ satisfy the usual commutation relations
\beq
[a^{\phantom{\star}}_{\mathbf{k}_l}, a^{\star}_{{\mathbf{k}_l}'}] = \delta_{{\mathbf{k}_l}{\mathbf{k}_l}'}.
\eeq
Comparison between the creation and annihilation operators of the \emph{in} and \emph{out} states will allow us to calculate the Bogoliubov coefficients relating one set to the other, thus leading us to evaluate the average number of emitted quanta.

\subsection{Greybody factor}\label{sec:greybody}

In our four dimensional problem, {\sl a priori} we cannot neglect the effect of
backscattering, i.e., the fact that, once emitted near the horizon,
photons can undergo with a certain probability, reflection back
in to the horizon due to the presence of a non-vanishing potential that
they encounter in their propagation \cite{haw-cmp}. This originates the so-called
greybody factor, which can be calculated as the square modulus of the
transmission coefficient. \\

Consider a massless scalar field propagating in a dielectric medium and
satisfying the massless Klein-Gordon equation in the pulse metric (\ref{static-pulse}),
and let us call $g(x)$ the component $g_{\tau\tau}$ appearing in (\ref{pulse-lapse}).
We get
\beq
\frac{n^2(x)}{c^2} \frac{1}{g(x)} \partial_\tau^2 \Phi-n(x) \partial_x
\left( \frac{1}{n(x)} g(x) \partial_x \Phi \right) -\partial_y^2 \Phi
-\partial_z^2 \Phi=0.
\eeq
We consider solutions of the form
\beq
\Phi (\tau,x,y,z) = e^{i\omega \tau} \varphi (x) e^{-i k_y y -i k_z z}.
\eeq
As a consequence, we obtain an equation of the form
\beq
(p(x) \varphi')' + \omega^2 k(x) \varphi -q(x)\varphi=0,
\eeq
where the prime stays for the derivative w.r.t. $x$ and
\beq
p(x) := \frac{g}{n};\quad k(x):= \frac{n}{c^2 g}; \quad q(x):= k_\perp^2
\frac{1}{n}.
\eeq
The following change of variable:
\beq
s(x):=\int^x du \sqrt{\frac{k(u)}{p(u)}}=\int^x du \frac{n}{c g},
\eeq
where $s$ results to coincide with the tortoise coordinate, leads to the
following Schrodinger-like form of the equation for $\varphi$:
\beq
\frac{d^2 \varphi}{d s^2} + (\omega^2 - Q(s)) \varphi =0,
\eeq
where
\beq
Q(s):=k_\perp^2 \frac{c^2 g}{n}.
\eeq
It can be noticed that, as $x\to x_+$, i.e. as $s\to -\infty$, one
obtains $Q(s)\to 0$, and that
\beq
\lim_{s\to \infty} Q(s) = k_\perp^2 \frac{c^2}{n_0^2}
\left(1-n_0^2 \frac{v^2}{c^2} \right) =: Q_{\infty}.
\eeq
$Q(s)$ turns out to be monotonically increasing from $0$ to the aforementioned
constant value as $s\to \infty$, and the transition from the zero value and
the asymptotically constant value for a bump is very fast,
in such a way that, as a plausible approximation for the calculation
of the transmission coefficient, we replace the above
potential $Q(s)$ with a step-like effective potential with height $Q_\infty$:
\beq
\frac{d^2 \varphi}{d s^2} + (\omega^2 - Q_{\infty}) \varphi =0.
\eeq
In this approximation, we can refer to well-known results of standard
quantum mechanics (see e.g. \cite{flugge},
where the step barrier appears as a subcase of problem 37), in order to
infer that:\\
a) the transmission coefficient is zero if
\beq
\omega < \sqrt{Q_{\infty}}=  k_\perp \frac{c}{n_0}
\sqrt{\left(1-n_0^2 \frac{v^2}{c^2}\right)}.
\eeq
b) the transmission coefficient is
\beq
|T|^2 = 4 \frac{\omega \sqrt{Q_{\infty}}}{(\omega+\sqrt{Q_{\infty}})^2} = \Gamma,
\eeq
for
\beq
\omega>\sqrt{Q_{\infty}}.
\eeq
We point out that the dispersion relation
\beq
k_x^2+k_\perp^2 = \frac{n_0^2}{c^2} \omega^2
\eeq
implies that a) cannot occur. As a consequence, the Heaviside function which, in principle,
should multiply the aforementioned $|T|^2$, is always equal to 1. 	

We have to translate the latter result in the laboratory frame;
the simple substitution
\beq
\omega = (\omega_l -v k_{xl}) \gamma,
\eeq
where the label $l$ indicates laboratory frame quantities, leads to
the following expression for the greybody factor:
\beq
\Gamma (\omega_l,{{\mathbf{k}}_\perp}) \simeq
\frac{4(\omega_l -v k_{xl}) \sqrt{(\omega_l -v k_{xl})^2-k_\perp^2 q_0^2}}
{\left[(\omega_l -v k_{xl})+k_\perp q_0\right]^2},
\eeq
where
\beq
q_0 = \frac{c}{n_0} \sqrt{1-n_0^2 \frac{v^2}{c^2}}.
\eeq
One can also take into account that
\beq
\omega_l -v k_{xl} = \omega_l \left(1-n_0 \frac{v}{c}\cos(\theta)\right),
\eeq
so that
\beqnl
\Gamma  &\simeq &
\frac{4(1-n_0 \frac{v}{c}\cos(\theta)) |n_0 \frac{v}{c}-\cos(\theta)|}
{\left[1-n_0 \frac{v}{c}\cos(\theta)+|n_0 \frac{v}{c}-\cos(\theta)|\right]^2}\cr
&\times& \theta_H (\frac{\omega}{q_0} (1-n_0 \frac{v}{c}\cos(\theta))-k_\perp).
\eeqnl
Being $\eta\ll 1$, and $v/c = (n_0+k\eta)^{-1}$, $k\in (0,1)$,
we obtain
\beqnl
&&1-n_0 \frac{v}{c}\cos(\theta) \sim 1-\cos(\theta)+\frac{k\eta}{n_0} \cos(\theta),\\
&&n_0 \frac{v}{c}-\cos(\theta) \sim 1-\cos(\theta)-\frac{k\eta}{n_0},
\eeqnl
which leads to the conclusion that
\beq
\Gamma \simeq  1. 
\eeq
It is worth pointing out that, on the grounds of the previous result,
a $90$-degrees emission would not suffer any substantial suppression by backscattering. This is particularly relevant in connection to recent measurements of 90-degree photon emission from RIP-induced horizon \cite{PRL}.

\subsection{Thermal spectrum}\label{sec:thermalspectrum}
In analogy with the corresponding black hole calculations, in order to find the evaporation temperature of the RIP we must
compute the expectation value of the number operator of the outgoing photons evaluated in the \emph{in} vacuum.
In this connection, it is important to remark that, contrary to the standard black hole scenario, in the case of
a RIP propagating in a nonlinear Kerr medium we have, strictly speaking, no real collapse situation. Nevertheless, we
can ``simulate a collapse'' by building up a refractive index perturbation starting
from an initial situation in which no signal is present in the dielectric.

We perform a coordinate shift which moves the black hole horizon in $u=0$. A photon starting at $u<0$ is trapped an cannot
reach the front observer. As a consequence, states with $u<0$ cannot
be available to the front observer and he will be lead to consider in his description
only the $u>0$ part of the $F^+$ solution, thus multiplying the latter by a Heaviside function $\theta(u)$.
The Bogoliubov coefficients $\alpha_{{\mathbf{k}_l} {\mathbf{k}_l}'}$ and $\beta_{{\mathbf{k}_l} {\mathbf{k}_l}'}$ relating, respectively, the positive and negative frequency components between the initial \emph{in} and final \emph{out} states are analysed in  ~\ref{B}, and satisfy the fundamental relation
\beq\label{relaz}
\sum_{{\mathbf{k}_l}'} \abs{\alpha_{{\mathbf{k}_l} {\mathbf{k}_l}'}}^2 = e^{\frac{4\pi c k_w}{n'(u_+) v}} \sum_{{\mathbf{k}_l}'} \abs{\beta_{{\mathbf{k}_l} {\mathbf{k}_l}'}}^2,
\eeq
which is also proved in the appendix.
Moreover, the expectation value of the number operator of the outgoing photons in the initial vacuum state is
\beq
\langle 0\; in | N_{{\mathbf{k}_l}}^{out} | 0\; in \rangle = \sum_{{\mathbf{k}_l}'} |\beta_{{\mathbf{k}_l} {\mathbf{k}_l}'}|^2,
\eeq
Then, by using the completeness relation for the Bogoliubov coefficients and taking into account
backscattering:
\beq
\sum_{{\mathbf{k}_l}'} \left(\abs{\alpha_{{\mathbf{k}_l}{\mathbf{k}_l}'}}^2 - \abs{\beta_{{\mathbf{k}_l}{\mathbf{k}_l}'}}^2 \right) = \Gamma (\omega_l,{{\mathbf{k}}_\perp}),
\eeq
and Eq.~\eqref{relaz}, we obtain the following thermal-like distribution written in terms of asymptotic (physical) frequencies:
\beq
\langle N_{{\mathbf{k}_l}}\rangle = \frac{\Gamma}{\exp\left( \frac{\hbar \omega_l}{k_b T}\right)-1}
\eeq
where
\beq
T = v^2\cfrac{\hbar}{2\pi k_B c}\ \cfrac{1}{1-\frac{v}{c}n_0 \cos{\theta}}\left|\cfrac{dn}{du}(u_+)\right|.
\eeq
By comparing this temperature with the temperature $T_+$ derived in the analogue Hawking model in the pulse frame (see Eq.~(\ref{Hawking_temp})), 
we find the relation:
\beq \label{temp1}
T = \frac{1}{\gamma} \frac{1}{1-\frac{v}{c}n_0\cos (\theta)} T_+.
\eeq	

For the same values of the parameters leading to Eq.~\eqref{t-pulse}, for $\theta=0$
and for a typical value $v\sim \frac 23 c$, we would obtain
\beq
T \sim 78 K,
\eeq
which is again much greater than the values of $T$ for typical acoustic black holes.
Moreover, in the specific case of a shock front, this temperature may increase significantly so
that $T\sim 2000 K$ (cf. subsection \ref{sec-shock}).

Equation.~\eqref{temp1} gives the correct transformation law for the temperature in going over from the pulse frame to the laboratory frame. Indeed, to find how the temperature transforms, start from Wien's displacement law which gives the wavelength of maximum emission of a black body as a function of the temperature
\beq
\lambda_\text{max} T = b = 2.9 \times 10^{-3} m \times K.
\eeq
Converting to the frequency of maximum emission gives
\beq\label{wienfreq}
\omega_{l \text{max}} = \frac{2\pi c}{b} T.
\eeq
Now, under the boost connecting the lab frame to the pulse frame the frequency transforms according to the relativistic Doppler formula in a medium with refractive index $n_0$
\beq\label{doppler}
\omega = \omega_l\gamma \left(1 - \frac{v}{c}n_0 \cos\theta\right),
\eeq
where $\omega$ is the frequency in the pulse frame, and $\theta$ is the emission angle with respect to the direction of motion of the pulse in the lab frame.
Combining Eq.~\eqref{doppler} with Eq.~\eqref{wienfreq} gives Eq.~\eqref{temp1}.

\begin{widetext}
\begin{center}
\begin{table}
\begin{tabular}{|c|c|}
\hline
&\\
{\sl Black Hole - Pulse Frame} & {\sl Quantum Field Theory - Lab Frame}\\
&\\
\hline
&\\
propagation equation (eikonal approximation) & propagation equation  (eikonal approximation)\\
$\downarrow$ & $\downarrow$\\
appearance of a bh horizon & onset of the surface s.t. $1-n(u) \frac vc =0$\\
$\downarrow$ & $\downarrow$\\
logarithmic singularity in the phase of quantum modes & logarithmic singularity in the phase of quantum modes\\
$\downarrow$ & $\downarrow$\\
thermal spectrum with temperature $T_+$ & thermal spectrum with temperature $T$\\
&\\
\hline
\end{tabular}
\caption{\label{table} Comparison between the analogue Hawking and QFT descriptions of quantum vacuum emission by a moving RIP.}
\end{table}
\end{center}
\end{widetext}

\section{Effects of optical dispersion}\label{sec:comment}

In the framework of analogue models for black holes (dumb holes, bec models) there is a
number of studies devoted to the analysis of the actual generation of Hawking radiation
(see e.g. \cite{unruh-disp,corley,jacobson,macher-bhwh,macher-bec,balbinot,corley-laser,
coutant-laser,finazzi,leonhardt}).
In particular, a suitable dispersion law, which
involves the fourth power of the momentum $k$ in the comoving frame is considered, since the
original model by Unruh \cite{unruh-disp}. Calculations, in particular in presence of a black hole - white hole
system,
are quite involved and require also numerical simulations.\\
We do not deal herein quantitatively with the problem of mode conversion,
and we do not consider the problem of black hole lasers
(see \cite{unruh-disp,corley,corley-laser,
jacobson,macher-bhwh,macher-bec,carusotto,coutant-laser,finazzi,balbinot,leonhardt}),
which will be taken into account in  further developments of the present model.
Still, we aim to point out the main physical consequences  due to
optical dispersion, and we shall stress that optical dispersion behaves
as a fundamental ingredient for dielectric models. In this respect, there is
agreement with dispersive models in sonic black holes.\\

We start by recalling how optical dispersion affects the
dispersion relation for the quantum electromagnetic field in linear homogeneous dielectric media
(see e.g.  \cite{barnett}):
\beq
n^2 (\omega_l)\; \omega_l^2 -{\mathbf{k}_l}^2 c^2 =0.
\label{hutt-barn}
\eeq
This dispersion relation is obeyed by the monochromatic components
of the field, and  in \cite{barnett} phenomenological quantization of the
electromagnetic field in a dielectric medium is justified on the grounds of a more rigorous
approach. In the presence of optical dispersion
\beq
\Phi (\mathbf{x}_l,t_l) = \int_0^{\infty} d\omega_l \phi (\omega_l,\mathbf{x}_l,t_l),
\eeq
where $\phi (\omega_l,\mathbf{x}_l,t_l)=\varphi_+ (\omega_l,\mathbf{x}_l) \hbox{e}^{-i\omega_l t_l} +
\varphi_- (\omega_l,\mathbf{x}_l) \hbox{e}^{i\omega_l t_l}$ are the monochromatic components.
Then it is easy to show that, even by allowing a spatial dependence for the refractive index
\beq
\nabla \varphi (\omega_l,\mathbf{x}_l)+\omega_l^2 \frac{n^2(\omega_l,\mathbf{x}_l)}{c^2} \varphi (\omega_l,\mathbf{x}_l)=0.
\label{mono-eq}
\eeq
Moreover, in the case of an homogeneous medium, by letting
$\varphi (\omega,\mathbf{x})\propto \exp( i {\mathbf{k}_l}\cdot \mathbf{x})$, one finds that
eqn. (\ref{mono-eq}) is satisfied only if
\beq
|{\mathbf{k}_l}|=\omega_l \frac{n(\omega_l)}{c},
\eeq
i.e. only if each monochromatic component travels at its phase velocity (as obvious).

By keeping into account a so-called 1-resonance model, we use the Sellmeier equation
\beq
n^2 (\omega_l) = 1 + \frac{\omega_c^2}{\omega_0^2-\omega_l^2},
\label{sellm}
\eeq
where $\omega_0$ is the resonance frequency and $\omega_c$ is the coupling constant, which is also
called plasma frequency. It is worth noting that the function $n^2 (\omega_l)$ is a Lorentz scalar
and it is invariant under boosts.
(\ref{sellm}) amounts to a quartic equation in $\omega_l$:
\beq
\omega_l^4 -(\omega_L^2+{\mathbf{k}_l}^2 c^2)\omega_l^2 +\omega_0^2 {\mathbf{k}_l}^2 c^2 =0,
\eeq
where $\omega_L^2\equiv \omega_0^2+\omega_c^2$ is the longitudinal resonance frequency.
Solutions are
\beq
\omega^2_{l\; \pm} = \frac 12 \left[\omega_L^2+{\mathbf{k}_l}^2 c^2
\pm \sqrt{(\omega_L^2+{\mathbf{k}_l}^2 c^2 )^2-4\omega_0^2
{\mathbf{k}_l}^2 c^2}\right].
\eeq
Then one finds two branches, which occur for
$0\leq \omega_{l-}<\omega_0$ and $\omega_L \leq \omega_{l+} <\infty$ respectively,
with a forbidden region in the interval $(\omega_0,\omega_L)$.\\
It is worth pointing out that, even if more complete expressions can be provided for the dispersion relation,
for practical purposes it is very useful the so called
Cauchy formula
\beq
n (\omega_l) = n_0 + B_0 \omega_l^2,
\label{cauchy}
\eeq
where $n_0$ is the would-be refractive index in absence of optical dispersion and $B_0$ is a suitable constant; this approximation works well for fused silica
in the visible frequency interval, and can also be obtained from the more complete
formula (\ref{sellm}) in the formal limit as $\omega_l\to 0$ (i.e., from a physical point of view,
as $\omega_l \ll \omega_0$).
For our aims, {\sl a posteriori}, its limited bandwidth validity is not a real problem, because
it turns out that also Hawking effect takes place in a limited frequency window.

\subsection{Dispersion and RIP}

\begin{figure}[t]
\includegraphics[width=8cm]{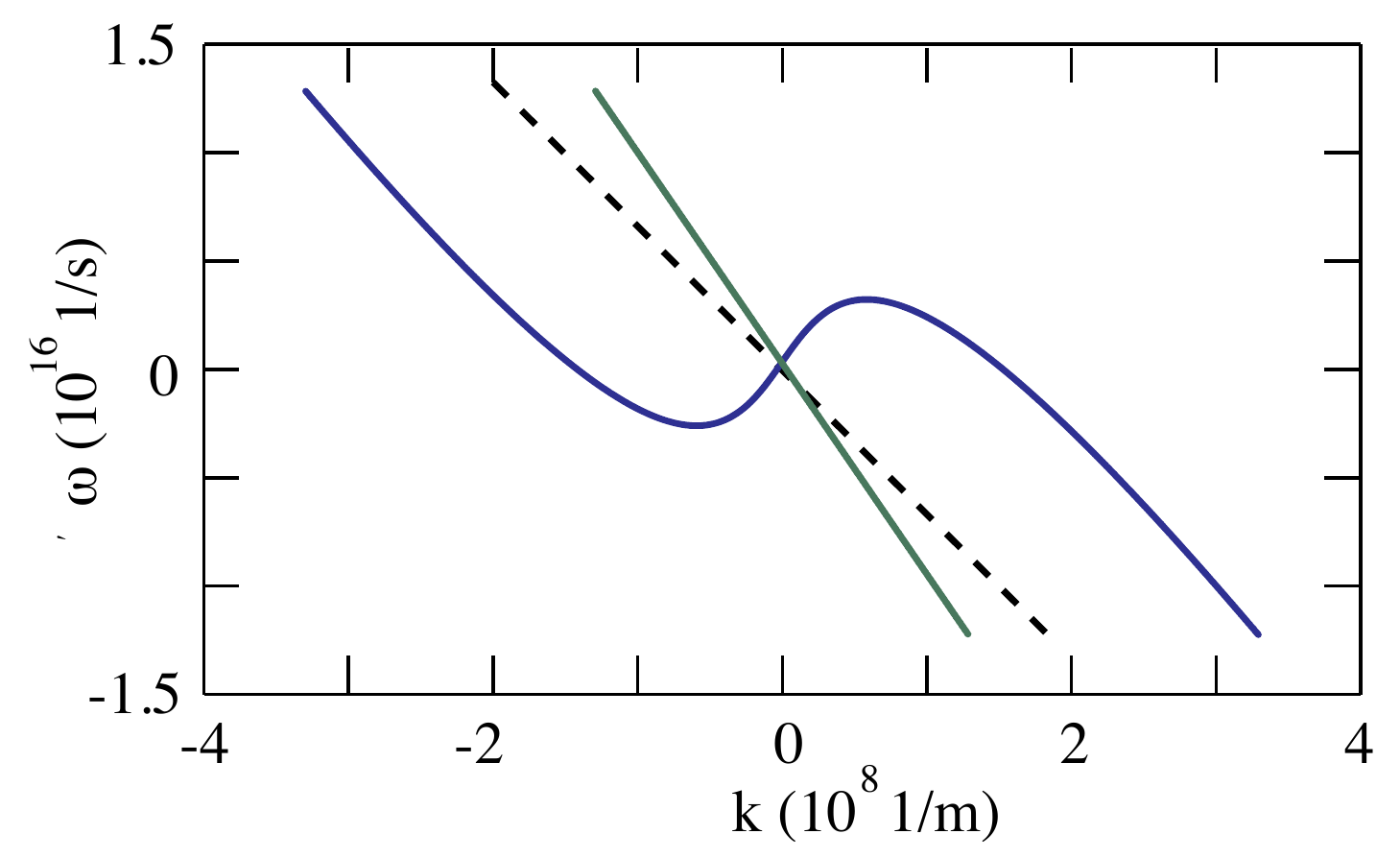}
\caption{ \label{fig:fig01} (in color online) Branches relative to the dispersion relation
(\ref{disp-cauchy}) in the comoving frame. The dashed line corresponds to $\omega = - v k_x$.}
\end{figure}

We  note that the dispersion relation for a linear homogeneous dielectric medium is quite different
from the one which has been hitherto considered in the sonic model by Unruh and its subsequent
developments (see also analogous comments in Ref. \cite{schutzhold}).\\
We may also wonder if the above relation can be trusted in our case, as we started from
a nonlinear medium which is also non-homogeneous because of the RIP.
We recall that we assume  to be involved with linearized
quantum fields around our effective geometry, and then it is reasonable to explore what happens in
presence of optical dispersion by considering linear dispersion effects. Moreover, we
make the following ansatz:
\beq
n(x_l-vt_l,\omega_l)=n_0 (\omega_l)+\eta\; f(x_l-vt_l),
\label{disp-ansatz}
\eeq
i.e. we neglect optical dispersion effects in the RIP, and keep trace of it only in
the background value $n_0$, which does not depend on spacetime variables, and
passes from the status of constant to that of a function depending on $\omega_l$.\\
In order to justify {\sl a posteriori} this ansatz
we can take into account what happens in the usual modelization of dielectric media when
inhomogeneities are considered. A possibility is to account for inhomogeneities by
including space-dependent density, polarizability, resonant frequencies, and so on
(see e.g. \cite{suttorp}). By neglecting dissipation in a polariton model, one can obtain a
dielectric susceptibility which depends on space and frequency, in such a way that the
Sellmeier formula changes only because of an explicit dependence of $\omega_c,\omega_0$ on
space variables. In our case, because of the properties of fused silica in the visible
region, where the Cauchy formula works well, we can even neglect the spatial dependence
of $\omega_0$ (which should be considered much greater than $\omega_l$ and $\omega_c$), and then,
recalling that $u:=x_l-v t_l$, on the grounds of Ref. \cite{suttorp} we
assume that
\beq
\omega^2_c (u)=\omega^2_{c0}+\delta\; \omega^2_{c1} (u),
\eeq
where 
$\delta\ll 1$ is a small parameter;
if we also define
\beq
n_0 (\omega_l)=\sqrt{1+\frac{\omega^2_{c0}}{\omega_0^2-\omega_l^2}},
\eeq
the Sellmeier equation (\ref{sellm}) then leads to
\beqnl
n (\omega_l,u) &=& n_0 (\omega_l) \sqrt{1+\delta\; \frac{1}{n^2_0 (\omega_l)}
\frac{\omega^2_{c1}(u)}{\omega_0^2-\omega_l^2}}\cr
&\sim& n_0 (\omega_l)+\frac{\delta}{2n_0 \omega_0^2} \omega^2_{c1}(u),
\eeqnl
where we have used the Cauchy approximation and neglected terms
order of $\delta \omega_l^2$. This validates our ansatz
(\ref{disp-ansatz}), with straightforward identifications.\\
Another possibility is to adopt in the visible frequency region the following perturbative ansatz:
\beqnl
n (\omega_l,u) &=& n_0 + \eta A_1 (u) + (B_0 +\eta  B_1 (u))\omega_l^2\cr
&\sim& n_0  + B_0 \omega_l^2 + \eta A_1 (u),
\eeqnl
where $n_0  + B_0 \omega_l^2 \equiv n_0 (\omega_l)$ and where we neglect terms $O(\eta \omega_l^2)$.
The latter approximation is very useful, because we can write
\beq
n (\omega_l,u) = n_0 (\omega_l) + \eta f (u) = n (u) + B_0 \omega_l^2,
\eeq
which isolates the contribution of the optical dispersion.\\
As for the dispersion relation, we obtain in the lab
\beq
n (\omega_l,u) \omega_l = \pm |{{\mathbf{k}_l}}| c,
\eeq
i.e.
\beq
n (u) \omega_l + B_0 \omega_l^3 \pm |{{\mathbf{k}_l}}| c=0.
\eeq
This can be carried into the comoving (pulse) frame,
where one obtains
\beqnl
&&n (x)\gamma ( \omega+v k_x) \pm \sqrt{\gamma^2
(k_x+ \frac{v}{c^2} \omega)^2 +k_\perp^2} c \cr
&&+ B_0 \gamma^3 ( \omega+v k_x)^3=0,
\label{disp-cauchy}
\eeqnl
i.e. the same dispersion relation as in absence of optical dispersion except for
the cubic term $\propto B_0$.
We shall consider in what follows the 2D reduction, where
$k_\perp=0$.
Graphical solutions of the above equation are displayed in figure \ref{fig:fig01}.
Note that
\beq
\omega_l = \gamma (\omega+v k_x),
\eeq
and then positive frequencies in the lab correspond to the region above
the dashed line
\beq
\omega = - v k_x.
\label{positive-lab}
\eeq

\subsection{Dispersion and horizons}

\begin{figure}[t]
\includegraphics[width=8cm]{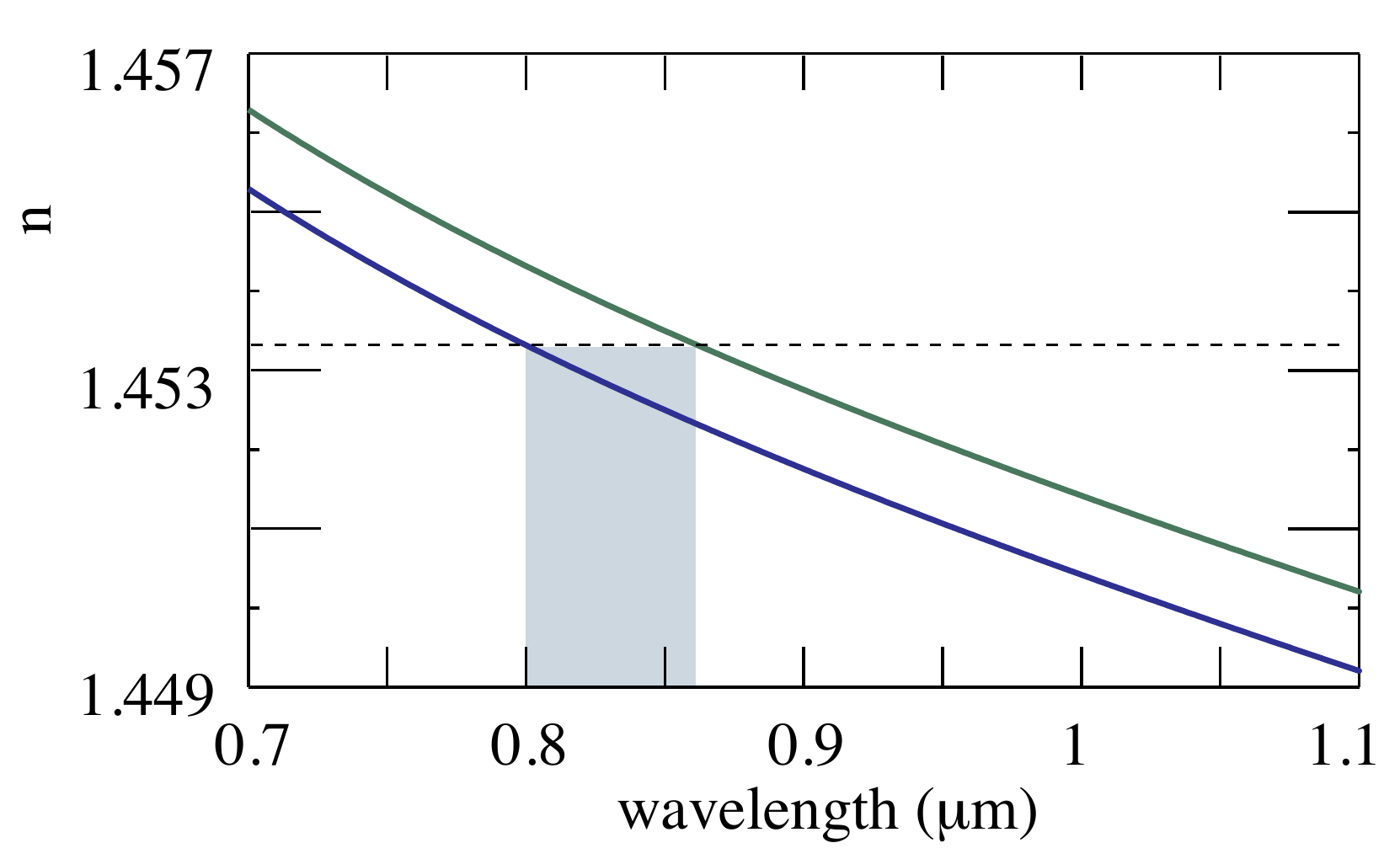}
\caption{ \label{fig:fig1} (in color online) Prediction of the Hawking emission spectral range accounting for the RIP velocity and the material refractive index $n$. The two curves show, for the case of fused silica, $n_0$ (in blue) and $n_0+\delta n$ (in green) with $\delta n=1\times10^{-3}$.  The shaded area indicates the spectral emission region predicted for a Bessel filament that has an effective refractive index $n=c/v_B$ where the Bessel peak velocity is $v_B=v_G/\cos\theta$ with cone angle $\theta=7$ deg and $v_G=d\omega/dk$ is the usual group velocity.}
\end{figure}

The definition of horizon which is suitable for this kind of situation, where
one cannot recover a metric by reinterpreting the dispersion relation as
an eikonal approximation of the wave equation (due to the presence of more than
quadratic terms), can {\sl a priori} involve phase velocity and/or group velocity;
cf. e.g. a short discussion in \cite{robertson}.
Let us first consider what happens when one involves the phase velocity,
indicated as $v_{l,\varphi}$ in the lab. reference frame and as $v_\varphi$ in the comoving frame:
\beq
v_{l,\varphi}=v \Leftrightarrow  v_\varphi =0,
\label{eq-phase}
\eeq
In the lab. frame the calculation is elementary, and leads to
\beq
v_{l,\varphi} = \pm \frac{c}{n (\omega_l,u)},
\eeq
which, due to (\ref{eq-phase}), implies the horizon condition
\beq
n (\omega_l,u) = \frac{c}{v};
\label{hor-cond}
\eeq
it is worth noticing that condition (\ref{hor-cond}) is a straightforward
generalization to the dispersive case of the condition obtained in the absence of
dispersion.
It is also important to point out  that the horizon condition in the lab in presence of
dispersion can be also expressed as follows:
\beq
1-n(\omega_l,u) \frac{v}{c}= 0,
\eeq
which e.g. in the Gaussian model provides
\beq
u_\pm = \pm \frac{\sigma}{\gamma} \sqrt{-2 \log \left( \left(\frac{c}{v}-n_0 (\omega_l)\right)
\frac{1}{\eta}\right)},
\eeq
and, in general, a modified horizon condition appears:
\beq
n_0 (\omega_l) < \frac{c}{v}\leq n_0 (\omega_l) +\eta.
\label{cond-disp}
\eeq
This condition is displayed in Fig.~\ref{fig:fig1} (where wavelength replaces $\omega_l$) and the velocity of the RIP is chosen to match the experimental conditions of Ref.~\cite{PRL}.
Explicitly, by using the Cauchy formula (\ref{cauchy}), we obtain (we consider only
positive frequencies) for $n_0 +\eta< \frac{c}{v}$, i.e. for a relatively small perturbation (see e.g. \cite{PRL})
\beq
\frac{\sqrt{\frac{c}{v}-n_0-\eta}}{\sqrt{B_0}} \leq \omega_l <
\frac{\sqrt{\frac{c}{v}-n_0}}{\sqrt{B_0}},
\label{cond-ph}
\eeq
i.e., in terms of wavelength,
\beq
\frac{2\pi c \sqrt{B_0}}{\sqrt{\frac{c}{v}-n_0}}
 < \lambda \leq
\frac{2\pi c \sqrt{B_0}}{\sqrt{\frac{c}{v}-n_0-\eta}}.
\eeq
If, instead, $n_0 +\eta\geq \frac{c}{v}$, i.e. for a substantially larger perturbation, we obtain
\beq
0 \leq \omega_l <
\frac{\sqrt{\frac{c}{v}-n_0}}{\sqrt{B_0}},
\label{cond-ph-eta}
\eeq
i.e., in terms of wavelength,
\beq
\frac{2\pi c \sqrt{B_0}}{\sqrt{\frac{c}{v}-n_0}}
 < \lambda < \infty
\eeq
For the parameters of Fig.~\ref{fig:fig1} (see caption), Eq. (\ref{cond-ph-eta}) is verified for $\eta>0.013$, a rather large yet not completely unrealistic refractive index variation.
We may evaluate the influence of material dispersion in relation to the event horizon condition.
In the absence of dispersion a horizon is created only if $v$ is tuned with extreme care such that
Eq.~(\ref{cond-disp}) with dispersionless $n_0$ is satisfied. Bearing in mind the typically small values of
$\delta n\sim 10^{-3}-10^{-4}$ (e.g. in fused silica, $n_2 \sim 3\cdot10^{-16}$ cm$^2$/W  and
$I\sim10^{13}$ W/cm$^2$), this would be no minor feat.

Note that the Hawking photons will
therefore in general be emitted only in a bounded spectral window. This is somewhat different from the
dispersion-less case in which, once $v$ is properly tuned so as to achieve the horizon condition, all
frequencies are simultaneously excited, and, {\sl a posteriori}, it can also justify why one can limit
oneself to adopt the Cauchy formula instead of a more complete Sellmeier.
The aforementioned finite spectral window for photon emission in turn implies that in the dispersion-less case one
should expect to observe the complete black-body spectrum predicted by Hawking. Conversely, in the
presence of dispersion in our analogue model, only a limited spectral region is excited and the
black-body spectral shape will not be discernible. Moreover, one has to take into account that,
even in homogeneous transparent dielectrics, optical dispersion affects the spectral
density of photons which is actually to be taken into account, by introducing a phase space factor
multiplying the standard Planckian distribution term which depends on both the refractive
index $n(\omega)$ and on the group velocity $v_g (\omega)$ \cite{milonni}:
\beq
\rho(\omega) = \frac{\left( \frac{\hbar \omega^3}{\pi^2 c^2}\right) \frac{n^2 (\omega)}{v_g (\omega)}}
{\exp \left(\frac{\hbar \omega}{k_b T}\right)-1}.
\eeq
This formula is intended to hold true in the reference of the thermal bath.
 These optical dispersion contributions
amounts to a sort of greybody factor arising from the interaction of photons with the dielectric
material. See e.g. \cite{milonni}. This is {\sl per se} a sufficient reason for
expecting deviations from the standard Planckian distribution. A very naive inclusion of dispersion in our case would also lead to $T=T(\omega)$, which would make even less plausible a pure
Planckian spectrum.

Regarding the width of the spectral emission window we note that this is determined by the value of
$\eta$. For a typical case we consider a Gaussian pump pulse in fused silica,
$n_g(800\textrm{ nm})=1.467$ and we find that for $\delta n=10^{-3}$, Eq.~(\ref{cond-disp}) is
satisfied over a $\sim$15 nm  bandwidth  at 435 nm. 
This window may become substantially large in lower dispersion regions as shown in  Fig.~\ref{fig:fig1}.

\begin{figure}[t]
\includegraphics[width=8cm]{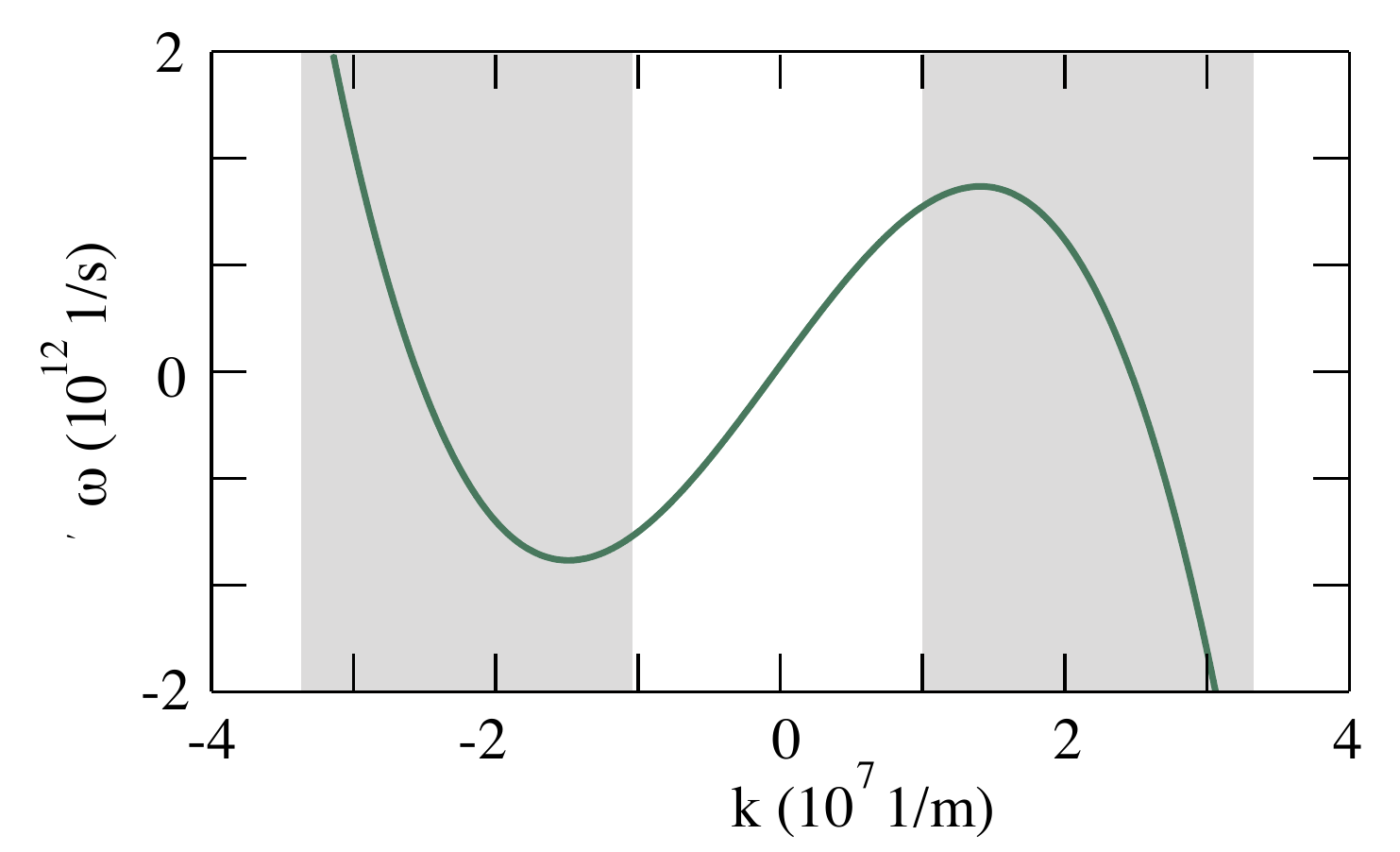}
\caption{ \label{fig:fig02} (in color online)
The dispersion relation
(96) in the comoving frame for a 2D model at a fixed point $x$ (only the branch relevant to horizon formation is shown). The shaded regions delimit the bands representing the values of $k_x$ for which a phase horizon  exists. See Eqs. (110-111). 
}
\end{figure}

The phase velocity horizon condition may also be calculated in the comoving frame, where one
obtains $v_{\varphi}=0$ if $\omega=0$ (for $k_x\not = 0$), i.e. for $v>0$
\beq
n \left(\gamma (\omega+v k_x),x\right)|_{\{\omega (k_x,x)=0,k_x\not =0\}} = \frac{c}{v}.
\eeq
Notice that $\omega=\omega (k_x,x)$ is solution of the 2D reduction of eqn. (\ref{disp-cauchy}).
The above condition requires that, as in absence of dispersion, $n_0<\frac{c}{v}$.
Moreover, if also $n_0+\eta<\frac{c}{v}$, then two disconnected regions are obtained:
\beqnl
-\frac{\sqrt{\frac{c}{v}-n_0}}{\gamma v \sqrt{B_0}} < &k_x& \leq -\frac{\sqrt{\frac{c}{v}
-n_0-\eta}}{\gamma v \sqrt{B_0}}\cr
\frac{\sqrt{\frac{c}{v}-n_0-\eta}}{\gamma v \sqrt{B_0}} \leq &k_x& <
\frac{\sqrt{\frac{c}{v}-n_0}}{\gamma v \sqrt{B_0}},
\label{cond-double}
\eeqnl
where only the latter region corresponds to $\omega_l>0$. These are shown in Fig.~\ref{fig:fig02}. Of course,
the latter condition is equivalent to condition (\ref{cond-ph}) in the lab frame.

If, instead,
$n_0+\eta\geq \frac{c}{v}$, then a unique connected region is obtained, with
\beq
-\frac{\sqrt{\frac{c}{v}-n_0}}{\gamma v \sqrt{B_0}} < k_x <
\frac{\sqrt{\frac{c}{v}-n_0}}{\gamma v \sqrt{B_0}}.
\label{cond-single}
\eeq
This condition is consistent with (\ref{cond-ph-eta}). For a further comment, see below.

One could also a priori consider a different definition of horizon, involving
the group velocity instead of the phase velocity:
\beq
v_{l,g}=v \Leftrightarrow  v_g =0,
\eeq
where the first equality holds in the lab and the second in the pulse frame.  For simplicity
we consider a 2D reduction of our model, so that $k_\perp=0$.  One obtains
\beq
v_{l,g}=v \Leftrightarrow \frac{c}{v}= n(\omega_l,u)+\frac{\partial n (\omega_l,u)}{\partial \omega_l}
\omega_l.
\eeq
In the case of the Cauchy formula, we find
\beq
\frac{c}{v}=n(\omega_l,u)+ 2 B_0 \omega_l^2\equiv n_g (\omega_l,u),
\label{hor-disp}
\eeq
where $n_g$ indicates the group refractive index.
In this case, in place of (\ref{cond-disp}), one would obtain
\beq
n_g (\omega_l) < \frac{c}{v}\leq n_g (\omega_l) +\eta,
\label{cond-disp-group}
\eeq
which presents a clear shift with respect to the window corresponding to (\ref{cond-disp}).
Explicitly, by referring again to (\ref{cauchy}), we find
\beq
\frac{\sqrt{\frac{c}{v}-n_0-\eta}}{\sqrt{3B_0}} \leq \omega_l <
\frac{\sqrt{\frac{c}{v}-n_0}}{\sqrt{3B_0}},
\label{cond-hp}
\eeq
i.e., in terms of wavelength,
\beq
\frac{2\pi c \sqrt{3B_0}}{\sqrt{\frac{c}{v}-n_0}}
 < \lambda \leq
\frac{2\pi c \sqrt{3B_0}}{\sqrt{\frac{c}{v}-n_0-\eta}}.
\eeq
It is easy to realize that coexistence of phase and group horizons is allowed only
if
\beq
\frac{\sqrt{\frac{c}{v}-n_0}}{\sqrt{3B_0}}\geq \frac{\sqrt{\frac{c}{v}-n_0-\eta}}{\sqrt{B_0}},
\eeq
i.e. if
\beq \label{overlap}
\eta \geq \frac{2}{3} \left(\frac{c}{v}-n_0\right).
\eeq

Note that experiments, at least
if the latter condition is not satisfied, should be able to distinguish between the two horizon conditions
for phase velocity and group velocity given above. For example, in the experimental conditions of Ref.~\cite{PRL} used also in Fig.~\ref{fig:fig1}, condition (\ref{overlap}) implies $\eta>0.009$. However the maximum $\eta$ obtained was only $\sim0.001$ and indeed the experiments clearly reveal emission in correspondence to the {\emph{phase}} horizon alone.

It is possible to show that, if $\eta> \frac{c}{v}-n_0$, there is a region in the $(\omega,k_x,x)$-space
where no group horizon appears because of
the lack of any real zero in the derivative of the dispersion equation.
In order to realize this fact analytically, we can proceed as follows. Let us
consider in the comoving frame the dispersion equation in the following form:
\beq
D=D_+ D_-=0,
\eeq
where
\beq
D_\pm : = n (x)\gamma ( \omega+v k_x) + B_0 \gamma^3 ( \omega+v k_x)^3
\pm \gamma
(k_x+ \frac{v}{c^2} \omega) c=0.
\eeq
The dispersion relation $\omega (k_x,x)$ solves $D=0$.
The only dispersion equation which can lead to a group velocity horizon is the
solution $\omega_- (k_x,x)$ of $D_-=0$. Instead, the other branch $\omega_+ (k_x,x)$ which solves
$D_+=0$ is monotonically decreasing.
We find
\beq
v_g=\frac{\partial \omega (k_x,x)}{\partial k_x}=\frac{\partial D}{\partial k_x}
\left( \frac{\partial D}{\partial \omega} \right)^{-1}.
\eeq
As a consequence, a group velocity horizon can emerge only by solving
the system $D_-=0$ and $\frac{\partial D_-}{\partial k_x}=0$. It is easy to show that
$\frac{\partial D_-}{\partial k_x}=0$ leads to
\beq
k_x^{(-)}= -\frac{\omega}{v}\pm \frac{1}{\gamma v \sqrt{3 B_0}} \sqrt{\frac{c}{v}-n(x)}.
\label{k-vgroup}
\eeq
It is immediate to see that, if $n_0+\eta\geq \frac{c}{v}$, as in the case where a unique
connected region is found for the phase velocity horizons (cf. eqn. (\ref{cond-single})),
$k_x^{(-)}$ in (\ref{k-vgroup}) is complex valued for all $x$ such that $n(x)>\frac{c}{v}$,
for which then group velocity horizons disappear.


We may also qualitatively describe the behavior of a monochromatic or quasi-monochromatic wave near a phase horizon. In the comoving frame, an horizon is approached only by waves that travel with
$v_\varphi>0$. Let us consider a wave traveling towards the white hole horizon; as it reaches the RIP,
the refractive index increases, and then it slows down. This effect is enhanced by dispersion,
because it implies a further increase of the refractive index. As the wave is as near as possible
to the phase (white hole) horizon, it nearly stops, but, as remarked by e.g. \cite{corley},
it cannot stop indefinitely, but, rather suffers reflection
(mode conversion). This happens for all frequencies belonging to the allowed windows indicated above.
The previous analysis is corroborated by the study of null geodesics, in the presence of optical
dispersion, which is carried out in \cite{cacciatori}. Indeed, therein it is shown (in a more general geometric
setting) that only geodesics with $v_{\varphi}\geq 0$ undergo a slowing down process but, instead of
suffering a process of trapping, as happens in absence of optical dispersion, they are reflected
away from the trapping point.\\
In the case of a wave packet, the dynamics appear to be more complex, but, due to its being a superposition
of monochromatic components, we can infer that what happens should be a sort of `remastering' of the
wave packet by the horizon in the following sense: frequencies beyond the allowed window are not
affected by the presence of the horizon, whereas frequencies in the aforementioned window
are selectively `bounced back' by a mode conversion mechanism. In this sense, we can appreciate
the peculiar behavior of the phase horizon as a sort of semipermeable membrane, or, even better,
as a sort of mirror with a selective reflection bandwidth,  which transmits
only the frequencies not belonging to the given windows.
The reflected frequencies will superimpose
maybe even in the form
of a wave packet. What is expected is that, in general, an incoming wave packet is converted into a reflected and
possibly very broad and irregular wave train.\\
The action of a group velocity horizon appears to be different, in the sense that, although the
existence of a window of allowed frequencies seems to be analogous to the one of a phase velocity
horizon, we have to point out that in the present case $\omega_l$ refers to the carrier wave in the
wave train. The group velocity horizon appears to be less selective than the phase velocity one:
it limits itself to cause a reflection of wave packets which have carrier frequencies in the
allowed window, without distinguishing between monochromatic components composing the packet itself.
The outgoing wave train is expected to be still in the form of a compact and relatively undisturbed wave packet.
Both numerical simulations and experimental results could be able to reveal this different
character of the two above horizon versions, and in particular measurements should be able to
discern which definition is relevant for the physics at hand. It is worth pointing out that,
in Ref. \cite{PRL}, the main role in photon production appears to be related to the phase
velocity horizon rather than to the group velocity horizon.


\section{Conclusions}\label{sec:conclusion}

We have studied the photon production induced in a dielectric medium by a refractive index perturbation,
which is created in a nonlinear medium by a laser pulse through the Kerr effect. The pulse has been assumed
to have constant velocity $v$. In the pulse frame, we have investigated the analogous metric, and we have
shown that two Killing horizons appear (black hole and white hole) for a generic but static
dielectric perturbation.
Despite the fact that the analogue metric is determined up to an overall conformal factor,
the temperature is conformally invariant. Then we have taken into account
quantum field theory in the lab frame, in a way that is independent of the
analogous geometric framework we have again shown that there is a photon production with a
thermal spectrum, corresponding to the Hawking effect. These findings are summarized in Table~\ref{table}. We have also provided the transformation
law of the temperature between the given frames and taken into account the effects of
dispersion on the horizon condition.\\

\acknowledgments
F.Belgiorno wishes to thank Sergio Serapioni and Lesaffre Italia S.p.A. for financial support
to the Department of Mathematics of the Universit\`a degli Studi di Milano, where part of
this work was performed.

\begin{appendix}

\section{Formal mapping to an acoustic black hole} \label{acou}

\begin{figure}
\centering
\includegraphics[width = 0.4\textwidth]{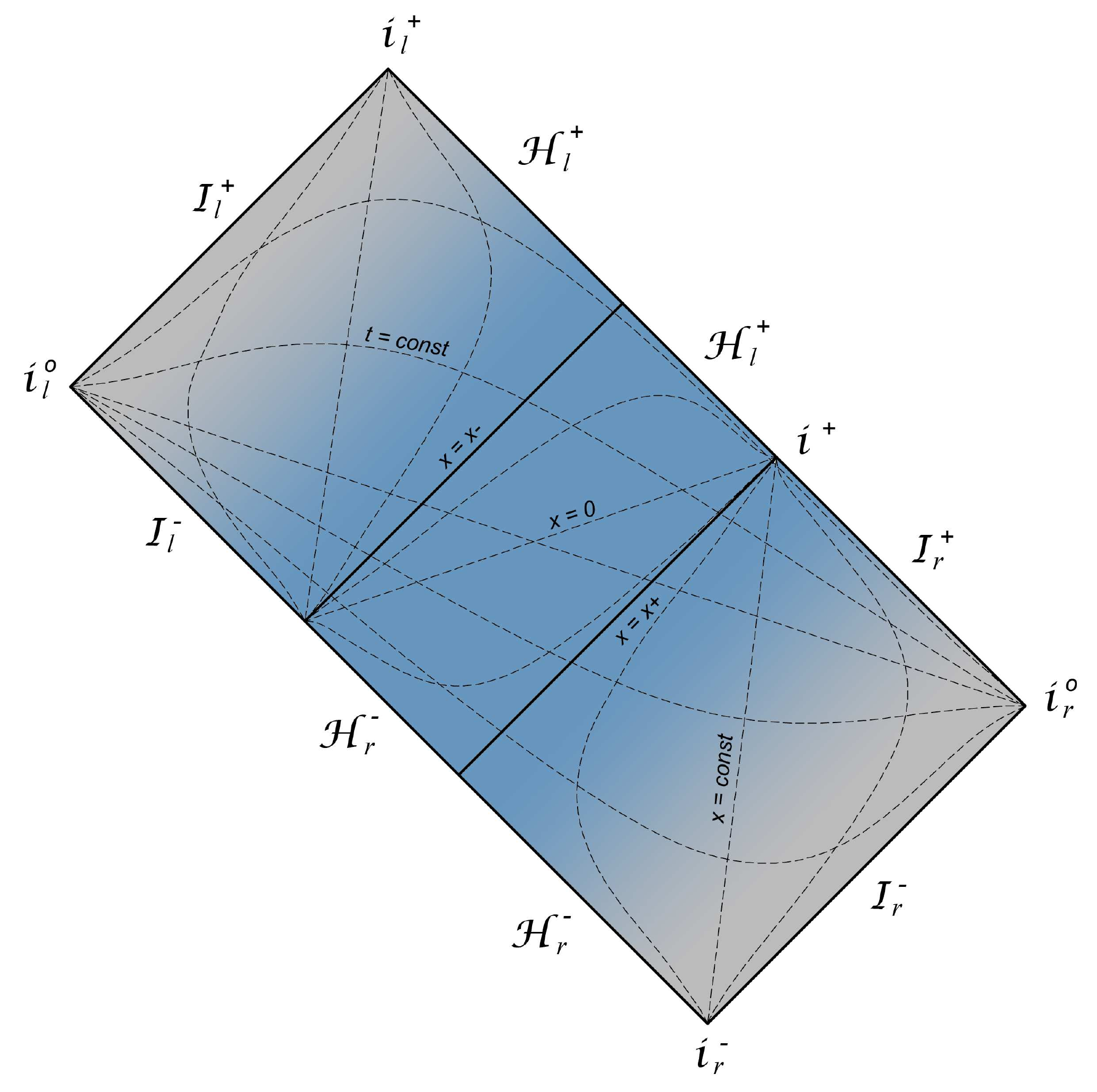}
\caption{Penrose diagram for the analogue metric~\eqref{metric}.}\label{fig3}
\end{figure}

It is interesting to notice that our black hole metric in the pulse frame can be mapped into a form
of acoustic black hole metric, provided that suitable identifications are made. In particular,
we refer to the acoustic black hole metrics taken into consideration in \cite{barcelo-causal},
and limit ourselves to the analysis of the $x-t$-part of the metric (indeed, only 2D metrics
are studied in \cite{barcelo-causal}). In particular, we are looking for a transformation allowing to
carry
\beqnl
ds^2_{(2)} &=& -\frac{\gamma^2}{n^2} \left[\vphantom{\frac{\gamma^2}{n^2}} -(c^2-n^2 v^2) dt^2 \right.\cr
 &-&\left. 2 v (1-n^2) dt dx + \left(n^2 -\frac{v^2}{c^2}\right) dx^2
\right]
\eeqnl
into the form
\beq
ds^2_{acoustic} = -\Omega^2 \left[ -({\tilde{c}}^2-{\tilde{v}}^2) dt^2 -2 \tilde{v} dt d\tilde{x} +  d\tilde{x}^2\right],
\eeq
where $\tilde{v}$ plays the role of fluid velocity (in general depending on $t,\tilde{x}$) and $\tilde{c}$ is the
local speed of sound, assumed to be constant as in \cite{barcelo-causal}. This re-mapping of the optical metric
into an acoustic one is implemented by the identifications
\beqnl
\tilde{c} & = & c,\\
\tilde{v} & = & \gamma^2 v \frac{n^2-1}{n},\\
\Omega^2  & = & \frac{1}{\gamma^2} \frac{1}{n^2 -\frac{v^2}{c^2}},\\
\frac{d\tilde{x}}{dx} & = & -\gamma^2 \frac{n^2 -\frac{v^2}{c^2}}{n}.
\eeqnl
It is straightforward to check that the horizon condition $\tilde{v}=c$ is equivalent to
the condition $n=\frac{c}{v}$, i.e. the solution of (\ref{horizons1D}).
By defining
\beqnl
d\tilde{u} = dt - \frac{d\tilde{x}}{\tilde{c}+\tilde{v}},\\
d\tilde{w} = dt + \frac{d\tilde{x}}{\tilde{c}-\tilde{v}},
\eeqnl
as in \cite{barcelo-causal}, one can also easily draw the Penrose diagram of our spacetime (see Fig.~\ref{fig3}), which is analogous to that of \cite[Fig.~28]{barcelo-causal}.
It is also straightforward to
verify that the surface gravity can be calculated also by means of the formula
\beq
\kappa_+ = \tilde{c} \left| \frac{\partial}{\partial \tilde{x}} (\tilde{c}-\tilde{v})\right|_{x=x_+}
\eeq
(cf. \cite{visser-acou}).

\section{the Hawking temperature and its conformal invariance}\label{temp-confo}

We consider the Euclidean version of the
metric and find that near the horizon the $ct-x$ part of the metric behaves like a cone
which becomes a flat plane only if a special choice of an angle parameter, to be related to the
inverse of the Hawking temperature, is chosen.

The following rescaling to dimensionless variables is performed:
\beqnl
c\tau \in (0,\beta) &\mapsto& \psi = \frac{2\pi}{\beta} c \tau \in (0,2\pi)\\
x \in (x_+,\infty) &\mapsto& \bar{x} = \frac{2\pi}{\beta} x \in \left(\frac{2\pi}{\beta} x_+,\infty\right),
\eeqnl
where $\tau$ stays for the Euclidean time.
Moreover, the following (local) diffeomorphism on the $ct-x$ part of the metric is
introduced:
\beqnl
ds^2_{(2)} &=&- \left(\frac{\beta}{2\pi}\right)^2
\left[ \frac{1}{n^2 (\bar{x})} g_{\tau \tau} (\bar{x} ) d (c^2 \tau^2) + \frac{1}{g_{\tau \tau} (\bar{x} )}
d{\bar{x}}^2\right]\cr
&=& - \left(\frac{\beta}{2\pi}\right)^2 H^2 (y)
\left( y^2 d\psi^2 + dy^2\right),
\eeqnl
whence
\beqnl
\frac{1}{n^2 (\bar{x})} g_{\tau \tau} (\bar{x}) &=& y^2 H^2 (y),\\
\frac{1}{g_{\tau \tau} (\bar{x})} \left( \frac{d\bar{x}}{dy} \right)^2 &=& H^2 (y).
\eeqnl
The above diffeomorphism is such that the $ct-x$ part of the metric near the horizon can, at least locally, be made conformal to a 2D plane with a suitable choice of $\beta$, with conformal factor
$\left(\frac{\beta}{2\pi}\right)^2 H^2 (y)$.
As a consequence, one finds (taking the  positive sign)
\beq
\frac{2\pi}{\beta}\frac{dx}{dy} = \frac{1}{y} \frac{1}{n(x)} g_{\tau \tau} (x),
\eeq
i.e.
\beq
y = A \exp \left(\frac{2\pi}{\beta}  \int^{x} dz n(z) \frac{1}{g_{\tau \tau} (z)} \right),
\eeq
where $A$ is an integration constant. The diffeomorphism  is defined to be regular
if we can include in the manifold also the point $y=0$, and then we have to require
\beq
\lim_{y\to 0^+} H^2 (y)= \lim_{y\to 0^+}  \frac{1}{y^2} g_{\tau \tau} (x(y)) = h_0^2,
\label{cond-conf}
\eeq
where $h_0^2$ is a finite positive constant. The above limit is equivalent to
\beq
\lim_{x\to x_+} \frac{1}{y^2 (x)} g_{\tau \tau} (x).
\label{cond-ess}
\eeq
Near the horizon one finds
\beq
y^2 \sim (x-x_+)^{\frac{2\pi}{\beta} \frac{c^2}{v^2} \frac{1}{\gamma^2}
\frac{1}{-\frac{dn}{dx}(x_+)}}
\label{y-beta}
\eeq
and, as a consequence, one has to choose
\beq
\beta = 2\pi \frac{c^2}{v^2} \frac{1}{\gamma^2}
\frac{1}{-\frac{dn}{dx}(x_+)} =: \hbar c\; \beta_h.
\eeq
The above method can be used also to confirm that the temperature does not depend
on the (static) conformal factor. 
Indeed, an overall conformal factor
$\Omega^2 (x)$ which is finite and non-vanishing at the horizons (
$\lim_{x\to x_{\pm}} \Omega^2 (x)= b^2_{\pm} >0$ ) does not modify Eq.~(\ref{y-beta}), as may be realized using Eqs.~(\ref{cond-conf}) and
(\ref{cond-ess}).


\section{Exact analysis of equation (\ref{eqnforA-4d}) near the singularity} \label{exact-pole}

We confine ourselves to discuss the root $u_+$, since all considerations can be trivially extended to $u_-$.
Then, we rewrite Eq.~\eqref{eqnforA-4d} as
\begin{equation}
A^{\prime\prime}(u) +  \frac{P(u)}{u-u_+} A^\prime(u) +\frac{Q(u)}{u-u_+}A(u) = 0,
\end{equation}
where we have introduced the functions $P(u)$ and $Q(u)$ (holomorphic in the disk $|u-u_+| < |u_+ - u_-|$)
\begin{gather}
P(u) = 2ik_w\frac{c^2+n^2(u)v^2}{c^2-n^2(u)v^2}, \\
Q(u) =  - \left(k_w^2+\frac{k_y^2+k_z^2}{1-n^2(u) \frac{v^2}{c^2}}\right).
\end{gather}
In a neighbourhood of $u_+$ Eq.~(\ref{eqnforA-4d}) writes, to leading order in $u-u_+$, as
\begin{multline}
A^{\prime\prime}(u) - \frac{2 i k_w c}{v n'(u_+)}\frac{1}{u-u_+}A'(u)+ \\ +\frac{(k_y^2+k_z^2)c}{2v n'(u_+) (u-u_+)} A(u) = 0,
\end{multline}
where the prime denotes derivation with respect to the lab variable $u$. The indicial equation is $\alpha(\alpha-1)- \alpha \frac{2 i k_w c}{v n'(u_+)} = 0$, with roots
$\alpha_1 = 0$ and $\alpha_2 = 1+\frac{2 i k_w c}{v n'(u_+)}$.
Therefore, in the neighbourhood of $u_+$, Eq.~(\ref{eqnforA-4d}) has two linearly independent solutions of the form
\begin{equation}
A^{(i)} (u) = (u-u_+)^{\alpha_i}\sum_{n=0}^{\infty} c_n^{(i)}(u-u_+)^n,
\end{equation}
$i=1,2$, where the series define holomorphic functions in the disc $|u-u_+|<|u_+-u_-|$ and whose coefficients can be obtained
recursively from the equation
$c_n^{(i)}(\alpha_i+n)(\alpha_i+n-1+p_0)+\sum_{r=0}^{n-1}((\alpha_i+r)p_{n-r}+q_{n-r-1})c^{(i)}_r= 0$,
where the $p_k$, $u_k$ ($k=0,1,2,\ldots$) are the coefficients of the expansion about $u_+$ of $P(u)$ and $Q(u)$ respectively.
Of particular interest is the solution (\ref{monochr}) corresponding to $\alpha=\alpha_2$, which for $u>u_+$
has the form
\beq
F_{\alpha_2} (u,w,y,z) = \xi (u) {\mathrm e}^{i\frac{2  k_w c}{v n'(u_+)}\log (u-u_+)+
i k_w w + i k_y y + i k_z z},
\eeq
where $\xi (u)$ is holomorphic in the neighbourhood of $u=u_+$ and vanishes as $u\to u_+$: indeed,
we have reabsorbed in it the factor $(u-u_+)$ associated with the real term in $\alpha_2$, in such a way that
\beq
\xi (u) = (u-u_+) \eta(u),
\eeq
where $\eta (u)$ is holomorphic in the neighbourhood of $u=u_+$ and
$\eta(u) = c_0+c_1 (u-u_+)+\ldots$ as $u\to u_+$.
We stress the presence of the logarithmic divergence of the phase as $u$ approaches $u_+$ even in this approach. Also, by comparison with the
study of the solutions contained in section III A, we can infer that the above solution
corresponds to the (exact) expansion of $F^+_{\mathbf{k}} (u,w,y,z)$ near $u=u_+$.

\section{Analytic continuations} \label{B}

We are interested in the Bogoliubov coefficient $\alpha_{\mathbf{k}\mathbf{k}'}$, which relates the positive frequencies between the initial \emph{in} state:
\begin{equation}
F^\text{in}_{\mathbf{k}} = e^{i k_x x + i k_y y + i k_z z - i\omega t },
\end{equation}
and final \emph{out} state $F_\mathbf{k}^+$:
\begin{equation}
F_\mathbf{k}^+ = \theta(u-u_+) \xi_{\mathbf{k}}(u) e^{i (\sigma \log(u-u_+) +k_w w + k_y y + k_z z)},
\end{equation}
where we have
introduced the shorthand notation $\sigma = \frac{2c}{n'(u_+) v}k_w$ and, in agreement with the
results obtained in Appendix \ref{exact-pole}, we have introduced also the analytic part $\xi (u)$.
It will be evident that, at least in the large frequency limit, this  $\xi (u)$ cannot affect
the thermal character of particle emission.

Let us start with the computation of $\alpha_{\mathbf{k}\mathbf{k}'}$, which, apart from a factor which will not affect our goal, which consists in the deduction of \eqref{eq-fin}, is given by
\beq\label{int1}
\alpha_{\mathbf{k}\mathbf{k}'} \propto \delta^2(\mathbf{k}_\perp-\mathbf{k}'_\perp)\int_0^{\infty}
\xi (u) u^{i\sigma} e^{-ik_u' u} du,
\eeq
where we have shifted the variable $u$ so that $u_+$ is mapped on $0$.
Since $2 k_u' = (k_x' + \omega'/v) > 0$, we see that the second exponential factor is rapidly decreasing at infinity when $\text{Im}(u)<0$. Consider the path $\Gamma$ starting from $0$ to $R>0$ along the real line, then following the arc of radius $R$ clockwise until $-iR$, and finally coming back from $-iR$ to $0$ along the imaginary axis. As the integrand is analytic inside the region bounded by the path,
the integral along $\Gamma$ vanishes for any positive value of $R$. Then, taking the limit $R\to +\infty$ we then see that the integral along the positive real axis is equal to the integral along the negative imaginary axis and
\beq\label{int}
\alpha_{\mathbf{k}\mathbf{k}'} \propto  e^{\frac{\pi}{2}\sigma}\frac{\delta^2(\mathbf{k}_\perp-\mathbf{k}'_\perp)}{i (k_u')^{1+i\sigma}}\int_0^\infty dt \, \xi\left(\frac{-i t}{k_u'}\right) t^{i\sigma} e^{-t}.
\eeq
We can consider the limit as $k'_u \gg 1$ and approximate $\xi$ for small values of its argument.
Then we obtain
\beq\label{alpha}
\alpha_{\mathbf{k}\mathbf{k}'} \propto   e^{\frac{\pi}{2}\sigma}\frac{\delta^2(\mathbf{k}_\perp-\mathbf{k}'_\perp)}{(k_u  ')^{2+i\sigma}}\Gamma(2 + i\sigma),
\eeq
where $\Gamma$ is the Euler Gamma function.

On the other hand, in order to calculate $\beta_{\mathbf{k}\mathbf{k}'}$, it is sufficient to revert the sign of $k_u'$  and $\mathbf{k}'$ in Eq.~\eqref{int1}.
This time, the integral can be calculated along the positive imaginary axis, since now the term relative to the integration on the arc vanishes for $\text{Im}(u) > 0$. Thus, rotating the path counter-clockwise, we obtain
\beq\label{beta}
\beta_{\mathbf{k}\mathbf{k}'} \propto   e^{-\frac{\pi}{2}\sigma}\frac{\delta^2(\mathbf{k}_\perp+\mathbf{k}'_\perp)}{(k_u')^{2+i\sigma}}\Gamma(2 + i\sigma),
\eeq
By comparing Eqs.~\eqref{alpha}  and~\eqref{beta}, it is easy to verify that
\beq\label{eq-fin}
\sum_{\mathbf{k}'} \abs{\alpha_{\mathbf{k}\mathbf{k}'}}^2 = e^{2\pi\sigma} \sum_{\mathbf{k}'} \abs{\beta_{\mathbf{k}\mathbf{k}'}}^2
\eeq
which is Eq.~\eqref{relaz}.

\end{appendix}


\end{document}